\documentclass[prb,aps,twocolumn]{revtex4}
\usepackage{epsfig,graphicx}

\begin{document}
\title{Spin-Hall effect and spin-Coulomb drag in doped semiconductors}
\author{E. M. Hankiewicz}
\address{Institut f\"ur Theoretische Physik und Astrophysik,
Universit\"at W\"urzburg, 97074 W\"urzburg, Germany}
\author{G. Vignale}
\address{Department of Physics and Astronomy, University of
Missouri, Columbia, Missouri 65211, USA}
\date{\today}
\begin{abstract}
In this review, we describe in detail two important spin-transport
phenomena: the extrinsic spin-Hall effect (coming from spin-orbit
interactions between electrons and impurities) and the
spin-Coulomb drag. The interplay of these two phenomena is
analyzed. In particular, we discuss the influence of scattering
between electrons with opposite spins on the spin current and the
spin accumulation produced by the spin-Hall effect. Future
challenges and open questions are briefly discussed.
\end{abstract}
\maketitle

\section {Introduction.}
In recent years two important spin-transport phenomena
\cite{Awschalom07} have been discovered in semiconductors, in the conducting (metallic) regime the spin-Hall effect (SHE)and
the spin-Coulomb drag.

The spin-Hall effect
\cite{Handbook,Murakami05r} is a bulk property of the semiconductor with a strong spin-orbit interactions
in the metallic regime. SHE is a close cousin of the anomalous
Hall effect. Anomalous Hall effect (AHE)
\cite{Luttinger54,Smit55,Smit58,Berger70a,Berger72,Lyo72,Nozieres,Bruno01,Jungwirth02,
Onoda02,Bruno05,Borunda07,Sinitsyn08} is the generation of a
transverse charge and spin polarization current in response to
an electric field. It appears in ferromagnets with strong
spin-orbit interactions like GaMnAs. In contrast, the spin-Hall
effect (SHE) is the generation of a transverse spin polarization
current alone in response to an electric field in a paramagnetic
medium with spin-orbit interactions and in the absence of a
magnetic field. By analogy with AHE there are two mechanisms
generating SHE: impurities and band structure. While the impurity
mechanism was suggested many years ago by Dyakonov and Perel
\cite{Dyakonov71,Perel,Hirsch99,Zhang00}, the second mechanism,
originating from band structure,  has come to the forefront in
recent years \cite{Murakami03,Sinova04}.  Again, similar to AHE, a
lively debate arose about which of these two mechanisms --
extrinsic (coming from impurities) or intrinsic (coming from band
structure)\cite{Culcer04,Loss04,Burkov04,Murak04,Murak041,Sinitsyn04,Inoue04,Shytov04,Schliemann05,Dimitrova05,Raimondi05,
Khaetskii06,Loss06,Gorini08} -- is more important, and how to distinguish
experimentally between the two \cite{Sarm06,Hankiewicz08}. Because of the transverse charge
response that comes with it, the AHE can be detected by purely
electrical measurements. However, this is not the case for the
SHE, because the spin-polarization current can not be directly
measured in transport. The spin-Hall effect in
semiconductors (GaAs, ZnSe) has  been mainly observed in optical
experiments \cite{Kato04,Sih05,Wunderlich05,Stern06,Stern08}. The idea of
these experiments is following: in a finite size sample, charge
current induces a transverse uniform gradient of spin density
(spin accumulation) which increases until the steady-state is
achieved. This spin accumulation can be measured quite clearly by
observing a change in polarization of a reflected beam of light
(Kerr effect).  This method has been successfully applied to n-type GaAs samples\cite{Kato04,Sih05,Stern06,Stern08}. In
another experiment performed on $p$-type GaAs, the
spin-accumulation was revealed by the polarization of the
recombination radiation of electrons and holes in a two-dimensional LED structure.
More recently, it has become possible to study the time evolution of optically injected charge and spin currents~\cite{Zhao06}, and to monitor the dynamics of spin accumulation in semiconductors~\cite{Stern08}.

The possibility of detecting the SHE by electrical measurements in
mesoscopic samples was theoretically suggested in
\cite{Hirsch99,Hankiewicz04}. In that proposal, an electric current driven
in one of the legs of an H-shaped  structure generates a
transverse spin current in the connecting part due to the SHE. Then, due to inverse
spin-Hall effect, this spin current produces a voltage difference
in the second leg of the structure \cite{Hankiewicz04}. Very recently this proposal was realized
in the H-shaped structures of the size of one micrometer, fabricated on the HgTe/CdTe quantum wells in the inverted regime \cite{Brune08}. These quantum wells characterize
a very long mean free path (larger than a couple of micrometer) and a very strong spin-orbit coupling \cite{Gui04} of the intrinsic (Rashba \cite{Rashba84}) type  and  authors concluded that observed voltage (of the order of microvolt) is the proof of the first measurement of ballistic spin-Hall effect in transport.
In a similar setup but using the
inverse spin-Hall effect alone \cite{Hankiewicz05}, the
spin-polarization was converted to electric signal and at least an order of
magnitude weaker electrical signal was
detected in metals, such as Al \cite{Tinkham06,Englert08}. The
origin of spin-Hall effect in metals is still under debate
\cite{Brataas05,Tanaka08,Murakami08}. Although theoretically
estimated intrinsic contribution in some of these materials (like
platinum) can be large \cite{Tanaka08,Murakami08}, the
calculations were performed, so far, for macroscopic systems and
did not include an extrinsic contribution. Therefore further
effort is warranted in this field, aimed at clear-cut
distinction between different mechanisms contributing to the
total spin-Hall effect in metals and semiconductors
\cite{Sarm06,Hankiewicz08}. In this review we mainly focus on the
extrinsic spin-Hall effect.

The quantum spin Hall (QSH) state is a novel topologically nontrivial
insulating state in semiconductors with strong spin-orbit interactions \cite{Kane05,ZhangQSHE06,ZhangSC06,Konig07,Konig08,Wu06,Xu06}, very different
from the SHE.  The QSH state, similar to the quantum Hall (QH) state, has a charge
excitation gap in the bulk. However, in contrast to the QH state, the QSH state does
not require existence of the magnetic field. Therefore for the QSH state,  time
reversal symmetry is not broken and instead of one spin degenerate edge channel (as
in the QH effect), two states with opposite spin-polarization counterpropagate at a
given edge. The QSH effect was first proposed by Kane and Mele for graphene \cite{Kane05}.
However, the gap opened by the spin-orbit interaction turned out to be extremely small
on the order of 10$^{-3}$ meV. Very recently Bernevig and Zhang theoretically proposed
that the QSH effect should be visible in HgTe/CdTe quantum wells with inverted band
structure \cite{ZhangSC06} and the experimental discovery of QSH effect in this material followed shortly afterwards
\cite{Konig07}.  QSH effect and the SHE effect are two distinct phenomena. While transport in the QSH effect occurs in the spin edge channels of an  insulating material, the SHE involves transport in the bulk of a conductor. This review is focused on the semiconductor spin transport in
the metallic regime, where the bulk is conducting. Specifically, we will summarize here the
current status of our knowledge concerning two important spin transport phenomena
in this regime: the spin-Hall effect and the spin-Coulomb drag. We refer a reader to
the recent review \cite{Konig08} for further details concerning the QSH effect.

The spin-Coulomb drag
\cite{Amico00,Flensberg01,Amico01,Amico02,Amico03,Vignale2007,Badalyan08}
is a many-body effect arising from the interaction between electrons with opposite spins,
which tends to suppress the relative motion of electrons with different spins and thus to reduce the spin diffusion constant.
This effect has been recently  observed in a (110) GaAs quantum well (which is essentially free of spin-orbit interaction)
by Weber {\it et al.}~\cite{Weber05} by monitoring the time evolution  of  a spatially varying pattern of
spin polarization, i.e. a spin grating.
The rate of decay of the amplitude grows in proportion to the square of the wave vector of the grating,
 and the coefficient of proportionality is just the spin diffusion constant.
 The measured value of the spin diffusion constant
 turns out to be much smaller than the single particle diffusion constant (deduced from the electrical mobility)
 and the difference can be quantitatively explained
 in terms of  Coulomb scattering between electrons of opposite spin orientation drifting in opposite direction,
 thus lending support to the theory of spin-Coulomb drag, as described in detail in Section III.

Intuitively when both the spin-Hall effect and  the spin-Coulomb drag
are present, the spin current generated by SHE should be reduced and
therefore it is important to take a look at the combined
influence of these two effects on spin transport.

The rest of the review is organized as
follows: in Section II we describe the extrinsic spin-Hall effect
using the Boltzmann equation approach; in Section III we
describe the spin-Coulomb drag effect; in Section IV we discuss in
detail the influence of spin-Coulomb drag on the extrinsic
spin-Hall effect; in Section V we briefly describe the intrinsic
spin-Hall effect and  the influence of intrinsic spin-orbit
coupling on the spin-Coulomb drag; in Section VI, we discuss
possible scenarios for the evolution of the SHE in semiconductors as a
function of mobility. Conclusions and open challenges are
presented in Section VII.

\section {Extrinsic spin-Hall effect.}
Fig.~(1) shows a setup for the measurement of the SHE. An electric
field (in the $x$ direction) is applied to a non-magnetic two
dimensional electron gas (2DEG). In response to this, a spin
current begins to flow in a direction perpendicular to the
electric field (the $y$ direction, see Fig.~(1)): that is to say,
spin up and spin down electrons with "up" and "down" defined in
respect  to the normal to the plane, drift in opposite directions
perpendicular to the electric field.
\begin{figure}[thb]
\vskip 0.27 in
\includegraphics[width=3.4in]{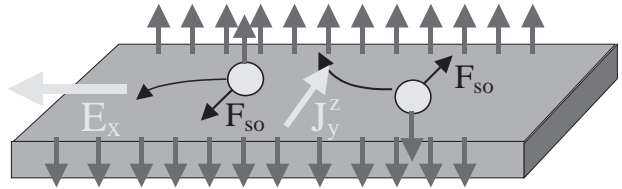}
\caption{Schematics of the spin-Hall effect. An
electric field applied  along the $x$ axis, $E_x$,  induces a
transverse spin current $J^{z}_{y}$.}
\end{figure}

There are two extrinsic mechanisms of generation of transverse
spin current: skew scattering and side jump
\cite{Zhang00,Engel05,Hankiewicz06}.
 They  both arise from the effect of the spin-orbit interaction on electron-impurity collisions.  Skew
-scattering  arises  from the asymmetry of the
electron-impurity scattering in the presence of spin-orbit
interactions: electrons  that are drifting in $+x$ direction under
the action of electric field are more likely to be scattered to
the left  than to the right if, say their spin is up, while the
reverse is true if their spin is down. This generates a net z-spin
current in the y direction. This mechanism is also known as ``Mott
scattering" \cite{Mott} and has been long known as a method to produce
spin-polarized beams of particles.

The second effect is  more
subtle and is caused  by the anomalous relationship between
the physical and canonical position operator, as will be explained
below. It is called ``side-jump", because semi-classically
it can be derived from a lateral shift in the position of wave packet during collision with impurities.
Without resorting to this description, we could arrive at the same side-jump term starting from the quantum kinetic equation and including the anomalous part of position operator. Figs. 2 and 3 present simple pictures of skew scattering
and side jump mechanisms respectively.
\begin{figure}[thb]
\vskip 0.27 in
\includegraphics[width=3.4in]{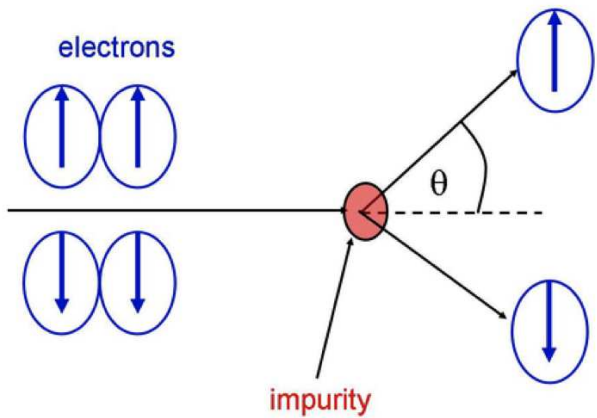}
\caption{Schematic picture of the skew-scattering effect. Due
to the spin-orbit interaction between electrons and impurities,
electrons with different spin orientations are deflected to opposite edges
of sample.}
\end{figure}

\begin{figure}[thb]
\vskip 0.27 in
\includegraphics[width=3.4in]{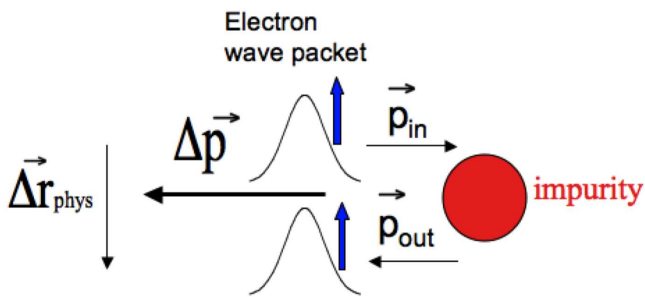}
\caption{Schematic picture of the side-jump in a head-on collision
with the impurity. The center of electron wave packet is shifted
by $\Delta \vec r_{phys}$ in a direction perpendicular to the
change of momentum during the collision. The physical origin of
this shift is described in the text.}
\end{figure}

After this brief, pictorial presentation of different mechanisms
contributing to the extrinsic spin-Hall effect, we are now ready to begin a more detailed analysis.

First of all, because both mechanisms depend crucially on the
spin-orbit interaction, it is necessary to say something about the
character of this interaction in the solid state.  The derivation
of this interaction involves steps that closely parallel the
derivation of  the spin-orbit interaction from the Dirac equation
for a single electron in vacuum. In that case, we arrive at the
effective one band (Pauli) Hamiltonian by applying a unitary
transformation\cite{Foldy50} that decouples the electrons from the
positrons, and then projecting onto the electron subspace.
However, during this transformation the position operator is
modified,  taking the form
\begin{equation}\label{r_phys}
\vec r_{phys,i} = \vec r_i-\alpha_0(\vec p_i \times \vec
\sigma_i)~,
\end{equation}
where $\alpha_0\hbar =\hbar^2/4m_e^2c^2 \approx -3.7 \times
10^{-6}$\AA$^2$ is the strength of spin-orbit interaction for bare
electron in vacuum and $m_e$ is the bare electron mass. This
physical position operator is valid for electrons in a conduction
band. For a general case, one has to apply the form of $r_{phys}$
as shown in Eq.~(74) of paper \cite{Sinitsyn08}. Substitution of
the modified position operator in the potential, followed by an
expansion to first order in $\alpha_0$,
 leads to the standard form of the spin-orbit interaction in vacuum:
\begin{equation}\label{H-so}
\hat H_{SO} = \alpha_0 (\vec{p} \times \vec \nabla V(\vec r))\cdot
\vec\sigma
\end{equation}
where $V(\vec r)$ is the external potential acting on an electron,
$\vec{p}$ is the electron momentum and $\vec{\sigma}$ is the vector
of the Pauli matrices.
In semiconductors, the spin-orbit (SO) interactions play double role.
First, we have SO effects induced by the periodic crystal potential $V_0(\vec r)$. This
causes splitting of the p-like valence band at $k=0$, in semiconductors like GaAs, into
a fourfold  degenerate band with total angular momentum $j=3/2$ (heavy and light hole bands)
and ``split-off band"  with $j=1/2$. Further, the periodic crystal field gives rise to a small spin-orbit interaction
of the order of $\frac{\hbar^2}{4m_e^2c^2}[\vec p \times \vec \nabla V_0(\vec(r))]\cdot \vec \sigma$ on electrons in the conduction band.  Second, there is the SO interaction induced by {\it any} external potential (different than $V_0(\vec r)$)if one wants to find an effective model
say for the conduction band. In other words, if we perform a similar unitary transformation (as we did for the Dirac model) in a semiconductor, folding eight bands (conduction band, heavy
and light hole bands as well as split-off valence band) into an
effective model for the conduction band then the resulting
spin-orbit interaction will be again of the same form as in
Eq.~(\ref{H-so}), but with a much larger ``coupling constant":
\cite{Winkler2003}
\begin{equation}\label{alpha}
\alpha=\frac{\hbar
P^2}{3m_e^2}\left[\frac{1}{E_g^2}-\frac{1}{(E_g+\Delta_{SO})^2}\right]~
\end{equation}
where $E_g$ is the gap energy between conduction and heavy/light
holes bands, $\Delta_{SO}$ is the splitting energy between
heavy/light holes and split-off bands, $P$ is the matrix element
of the momentum operator between the conduction and the
valence-band edges. Using values of the parameters appropriate for
the 2DEG in Al$_{0.1}$Ga$_{0.9}$As \cite{Sih05} with a conduction
band mass $m=0.074m_e$ we find $\alpha\hbar = 4.4\AA^2$. Therefore
in the conduction band of semiconductors the spin-orbit interaction is
six orders of magnitude larger than in vacuum and has the opposite
sign. Obviously the spin-orbit interaction induced by the periodic crystal field on the conduction band can be omitted as a correction many orders of magnitude smaller.
However, $\alpha\hbar$  is much smaller than the square of
the effective Bohr radius in GaAs (~$10^4\AA^2$) and in this sense
the spin-orbit coupling can still be considered a small
perturbation. Notice that the form of the physical position
operator for the conduction band is described by
Eq.~(\ref{r_phys}), with $\alpha_0$ replaced by  $\alpha$.

Taking this into account and omitting for the time being the
electron-electron interaction, we see that our effective
Hamiltonian for electrons in the conduction band of GaAs takes the
form
\begin{equation} \label{Hni}
H_{ni}=H_0+V_{ei}(\vec r_{phys})+E(\vec r_{phys}),
\end{equation}
where
\begin{equation}\label{H0}
H_0= \sum_i \frac{\vec {p_i}^2}{2m}
\end{equation}
is the kinetic energy of electrons in conduction band ($m$ being
the effective mass of the conduction band) and
\begin{equation}\label{Vei}
V_{ei}(\vec r_{phys})\simeq V_{ei}(\vec r_i) + \alpha \sum_i \left
[\vec p_i \times \vec \nabla_i V_{ei}(\vec r_i) \right ]\cdot \vec
\sigma_i
\end{equation}
where $V_{ei}(\vec r_{i})$ is the impurity potential and  the spin-orbit
interaction between electrons and impurities arises from the Taylor expansion of the impurity potential
around the canonical position operator $\vec r_i$. Finally,
\begin{equation}\label{HE}
E(\vec r_{phys})\simeq \sum_i \left \{ e \vec E \cdot \vec r_i+e
\alpha (\vec p_i \times \vec E) \cdot \vec \sigma_i\right\}
\end{equation}
where we took the interaction
with the external electric field $\vec E$  to be $e\vec E \cdot \vec r_{phys}$.

The physical  velocity operator is the time derivative of the
physical position operator i.e.:
\begin{eqnarray}\label{vphys}
\vec v_{phys}= -\frac{i}{\hbar}[\vec r_{phys},\hat H],
\end{eqnarray}
and has the form
\begin{eqnarray}\label{vphys1}
\vec v_{phys}= \frac{\vec p_i}{m}+ \alpha\left[\vec \nabla_i
V_{ei}({\vec r}_i) + e \vec E \right] \times \vec \sigma_i
-\alpha\frac{d\vec p_i}{dt}\times \sigma_i,
\end{eqnarray}
where the first two terms on the right hand side are derivatives
of the canonical position operator while the last term originates from
the time derivative of the anomalous part of position operator.
However, since the total force $\vec F_i = d\vec p_i/dt$ consists
of a force originating from impurities and one from the electric
field $F_i = -\vec \nabla_i V_{ei}({\vec r}_i) - e \vec E$, the
second and last terms of Eq.~(\ref{vphys1}) are equivalent.
Therefore $\vec v_{phys}$ can be written in the following compact
form:
\begin{eqnarray}\label{vphys2}
\vec v_{phys}= \frac{\vec p_i}{m}+ 2\alpha\left[\vec \nabla_i
V_{ei}({\vec r}_i) + e \vec E \right] \times \vec \sigma_i~,
\end{eqnarray}
One can see that in our model the z-component of spin is conserved
because it commutes with the Hamiltonian. We exploit the
conservation of $\sigma_z$ by defining the quasi-classical
one-particle distribution function $f_{\sigma}(\vec r,\vec k,t)$,
i.e. the probability of finding an electron with $z$-component of
the spin $S_z = \frac{\hbar}{2}\sigma$,  with $\sigma = \pm 1$, at
position $\vec r$ with momentum $\vec p = \hbar \vec k$ at the
time $t$. In this review we focus on spatially homogeneous
steady-state situations, in which $f_\sigma$ does not depend on
$\vec r$ and $t$. We write
\begin{equation}\label{fsigma}
f_{\sigma}(\vec r,\vec k,t) =
f_{0\sigma}(\epsilon_k)+f_{1\sigma}(\vec k)~,
\end{equation}
where $f_{0\sigma}(\epsilon_k)$ is the equilibrium distribution
function -- a function of the free particle energy $\epsilon_k
=\frac{\hbar^2 k^2}{2 m}$ --  and $f_{1\sigma}(\vec k)$ is a small
deviation from equilibrium induced by the application of  steady
electric fields $\vec E_\sigma$ ($\sigma=\pm 1$) which couple
independently  to each of the two spin components.  In the next
few sections we will apply the Boltzmann equation approach to
calculate $f_1$ taking spin Hall effect and spin Coulomb drag into
account on equal footing.\cite{Hankiewicz06}

To first order in $\vec E_\sigma$ the Boltzmann
equation takes the form
\begin{equation}\label{Boltzmann.Equation}
-e \vec E_\sigma \cdot \frac{\hbar \vec k}{m}
f^\prime_{0\sigma}(\epsilon_k) =\dot f_{1\sigma}(\vec k)_{c}~,
\end{equation}
where $f^\prime_{0\sigma}(\epsilon_k)$ is the first derivative of
equilibrium distribution function with respect to the energy and
$\dot f_{1\sigma}(\vec k)_{c}$ is the first-order in $\vec
E_\sigma$ part of the collisional time derivative $\dot
f_{\sigma}(\vec k)_{c}$ due to various scattering mechanisms. For the
electron-impurity scattering mechanism the collisional time derivative  has the following form:
\begin{eqnarray}\label{collision.integral}
\dot f_{\sigma}(\vec k)_{c,imp}  = -\sum_{\vec k'}\left[W_{\vec
k\vec k'\sigma}f_\sigma(\vec k) -W_{\vec k'\vec k\sigma}
f_\sigma(\vec k')\right] \delta(\widetilde
\epsilon_{k\sigma}-\widetilde \epsilon_{k'\sigma})\nonumber\\
\end{eqnarray}
where $W_{\vec k\vec k'\sigma}$ is the scattering rate for a
spin-$\sigma$ electron to go from $\vec k$ to $\vec k'$, and
$\widetilde \epsilon_{k \sigma}$ is the particle energy, including
an additional spin-orbit interaction energy due the electric field:
\begin{equation}\label{epsilon_tylda}
\widetilde{\epsilon}_{k\sigma}= \epsilon_k +2e\alpha\hbar\sigma
(\vec{E}_{\sigma}\times\hat{z})\cdot \vec{k}~,
\end{equation}
The peculiar form of  $\widetilde{\epsilon}_{k\sigma}$, which
differs from the naive expectation $\epsilon_k +e \vec E \cdot
\vec r_{phys}$ by a factor $2$ in the second term,  is absolutely
vital for a correct treatment of the ``side-jump" contribution.
The reason for the factor $2$ is  that the $\delta$ function in
Eq.~(\ref{collision.integral}) expresses the conservation of
energy in a scattering process. Scattering is a time-dependent
process: therefore the correct expression  for the change  in
position of the electron $\Delta r_{phys}$ must be calculated  as
the integral of the velocity over time:
\begin{equation}\label{Delta_r}
\Delta \vec r_{phys} =\int_{-\infty}^{\infty} \vec v_{phys}dt
\end{equation}
Before solving integral in Eq.~(\ref{Delta_r}), let's think for a
moment about scattering event. Let's take $\Delta \vec {p} = \vec
p_{out}-\vec p_{in}$ (see Fig.~(3)) to be the change in
momentum of an electron wave packet during collision with an
impurity. During the very short time of collision, $\nabla V(\vec
r) =-d\vec{p}/dt$ is very large and therefore second term of
Eq.~(\ref{vphys2}) completely dominates the velocity. Therefore,
we disregard first term in the velocity formula and obtain a
following form for the electron wave packet displacement:
\begin{equation}\label{Delta_r2}
\Delta \vec r_{phys} = -2\alpha \Delta\vec p\times\sigma /\hbar
\end{equation}
Therefore, Eqs.~(\ref{vphys1},\ref{epsilon_tylda},\ref{Delta_r})
are consistent.
\subsection{Skew-scattering}
From the general scattering theory, developed for instance in
\cite{Landau3}, one can deduce the form of scattering probability
from $\vec k$ to $\vec k'$ \cite{Kohn58,Hankiewicz06} as:
\begin{equation}\label{Wkk}
W_{\vec k\vec k',\sigma} = \left[W^s_{\vec k\vec k'}+\sigma
W^a_{\vec k\vec k'} (\hat k \times \hat k')_z\right]
\delta(\epsilon_k-\epsilon_{k'})~,
\end{equation}
where for centrally symmetric  scattering potentials $W^s_{\vec
k\vec k'}$ and $W^a_{\vec k\vec k'}$ depend on the magnitude of
vectors $\vec k$ and $\vec k'$ and the angle $\theta$ between
them. Furthermore the left/right asymmetry of skew scattering is
included explicitly in the factor $(\hat k \times \hat k')_z$ and
therefore both $W^s_{\vec k\vec k'}$ and $W^a_{\vec k\vec k'}$ are
symmetric under interchange of $\vec {k}$ and $\vec{k'}$. Taking
into account the form of the scattering probability as well as the
conservation energy during the scattering process we obtain
following expression for the linearized collisional derivative:
\begin{eqnarray}\label{collision.integral.2}
\dot f_{1\sigma}(\vec k)_{c,imp} = -\sum_{\vec k'} W^s_{\vec
k\vec k'} \left\{f_{1\sigma}(\vec k)-f_{1\sigma}(\vec
k')\right\}\delta(\epsilon_k-\epsilon_{k'})\nonumber\\ -\sigma
\sum_{\vec k'} W^a_{\vec k\vec k'} (\hat k \times \hat
k')_z\left\{f_{1\sigma}(\vec k)+ f_{1\sigma}(\vec
k')\right\}\delta(\epsilon_k-\epsilon_{k'})\nonumber\\ +2\sigma
\sum_{\vec k'} W^s_{\vec k\vec k'} f'_{0\sigma}(\epsilon_k)
e\alpha\hbar (\vec{E}_{\sigma}\times\hat{z})\cdot
(\vec{k}-\vec{k'})\delta(\epsilon_k-\epsilon_{k'})~,\nonumber\\
\end{eqnarray}
where $f_{1\sigma}(\vec k) = -f'_{0 \sigma}(\epsilon_k)\hbar\vec
k\cdot \vec V_{\sigma}(k)$ and  the first term on r.h.s. of this
formula is the symmetric scattering term, the second one is
the skew scattering term, while the last one will be ultimately
responsible for the side jump. To find the drift velocity, $\vec
V_{\sigma}(k)$, we need to multiply both sides of Boltzmann
equation~(\ref{Boltzmann.Equation}) by $\hbar \vec k/m$ and
integrate over $\vec k$ space, and therefore derive $\vec
V_{\sigma}(\vec k)$ self-consistently from the condition:
\begin{equation}\label{Boltzmann.Equation.2}
-e\sum_{\vec k} \frac{\hbar \vec k}{m}\left[\vec E_\sigma \cdot
\frac{\hbar \vec k}{m}\right]f^\prime_{0\sigma}(\epsilon_k)=
\sum_{\vec k} \frac{\hbar \vec k}{m} \dot f_{1\sigma}(\vec
k)_{c,imp}~.
\end{equation}
After substituting of collisional derivative
Eq.~(\ref{collision.integral.2}) into
Eq.~(\ref{Boltzmann.Equation.2}) and with the assumption that we
can omit the k-dependence of drift velocity in low temperatures,
we arrive at the following formula for $\vec V_\sigma$ to first
order in the spin-orbit interaction:
\begin{equation}\label{Vsigma}
\vec V_\sigma = \frac{-e\tau_\sigma}{m}\left[\vec E_\sigma -
\sigma \frac{\tau_\sigma}{\tau_\sigma^{ss}}
   \vec E_\sigma \times \hat z \right]-2e\alpha \sigma
(\vec{E_{\sigma}}\times \hat{z})~,
\end{equation}
where in the limit of zero temperature the symmetric scattering rate $1/\tau$ and
the skew scattering  rate $1/\tau_{ss}$ simplify to:
\begin{equation}\label{tau.regular.2}
\frac{1}{\tau_\sigma} \stackrel{T \to 0}{\simeq} \frac{m {\cal
A}}{4 \pi^2 \hbar^2}\int_0^{2\pi} d \theta~ W^s(k_F, \theta)
(1-\cos \theta)~,
\end{equation}
and
\begin{equation}\label{tau.anomalous.2}
\frac{1}{\tau^{ss}_\sigma} \stackrel{T \to 0}{\simeq}  \frac{m
{\cal A}}{4 \pi^2 \hbar^2}\int_0^{2\pi} d \theta~ W^a(k_F, \theta)
\sin^2 \theta~
\end{equation}.
\begin{figure}[thb]
\vskip 0.27 in
\includegraphics[width=3.4in]{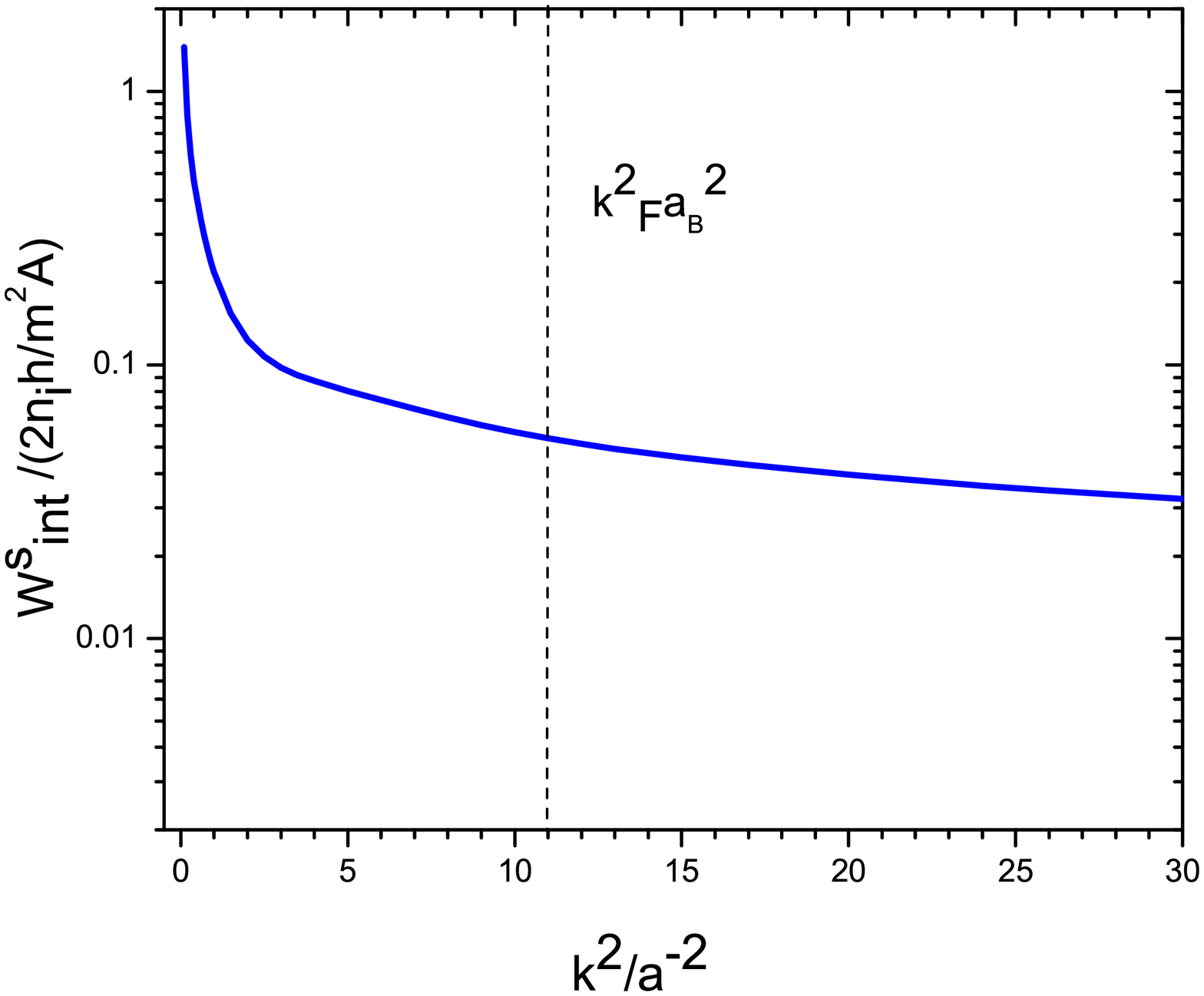}
\caption{Integrated symmetric scattering rate in units of $2n_i
h/m^2\cal A$ as a function of $k^2$ for a model circular well
attractive potential $V_0=-5$meV and radius $a=9.45$nm (described
in Reference \cite{Hankiewicz06}). We choose the parameters
typical for the experimental 2DEG
confined in Al$_{0.1}$Ga$_{0.9}$As quantum well i.e. 
density of electrons and impurities $n_{2D}=n_i=2.0\times
10^{12}$cm$^{-2}$, $m=0.074$m$_e$, and mobility $\mu
=0.1$m$^2/$V.s. The effective spin-orbit coupling  $\alpha\hbar=
4.4{\AA}^2$ in accordance with \cite{Winkler2003}.}
\end{figure}
\begin{figure}[thb]
\vskip 0.27 in
\includegraphics[width=3.4in]{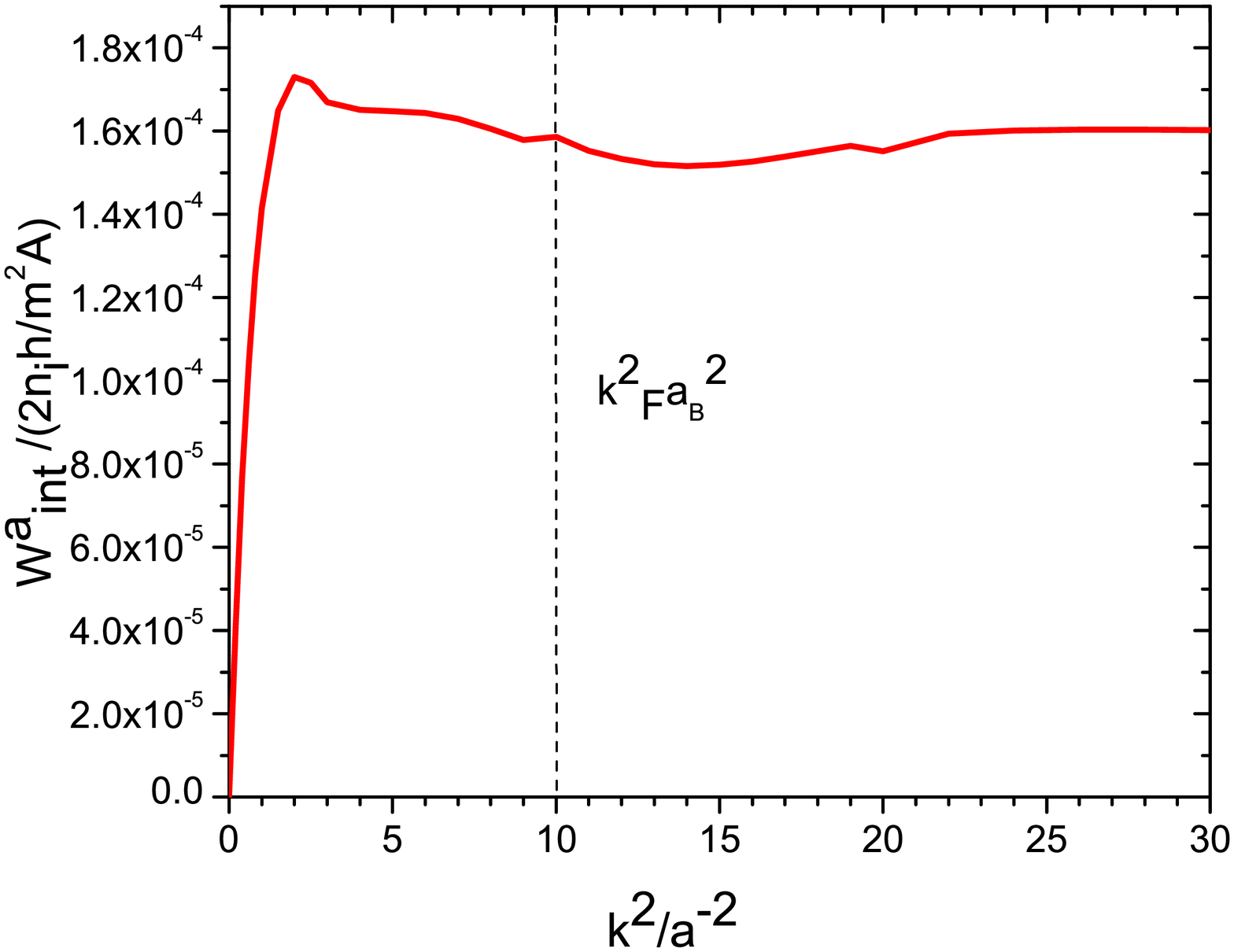}
\caption{Integrated asymmetric scattering rate in units of $2n_i
h/m^2\cal A$ as a function of $k^2$ for a model circular well
attractive potential $V_0=-5$meV and radius $a=9.45$nm (described
in Reference \cite{Hankiewicz05}). We choose the parameters
typical for the experimental 2DEG
confined in Al$_{0.1}$Ga$_{0.9}$As quantum well i.e. 
density of electrons and impurities $n_{2D}=n_i=2.0\times
10^{12}$cm$^{-2}$, $m=0.074$m$_e$, and mobility $\mu
=0.1$m$^2/$V.s. The effective spin-orbit coupling  $\alpha\hbar=
4.4{\AA}^2$ in accordance with \cite{Winkler2003}.}
\end{figure}
\begin{figure}[thb]
\includegraphics[width=2.0in]{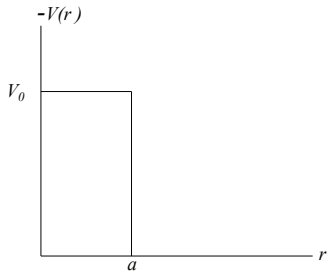}
\caption{The electron-impurity potential V($\vec r$) used to
estimate the ratio of symmetric to asymmetric scattering.}
\end{figure}
In Fig.~(4) and Fig.~(5), we present the integrated symmetric and
asymmetric scattering rates calculated from
Eqs.~(\ref{tau.regular.2}) and ~(\ref{tau.anomalous.2}) for a
simple step potential of the form:
\begin{equation}\label{scatter_V}
V(r)= V_0\theta(a-r)+ \bar{\alpha}aL_zS_z\delta(r-a)V_0 ~,
\end{equation}
where $V_0$ is the attractive electron-impurity potential,
$\bar{\alpha} = \alpha \hbar/a^2$ is the renormalized spin-orbit
interaction, $\alpha$ is the effective spin-orbit coupling
constant for the conduction band, $a$ is the impurity radius,
$L_z$ and $S_z$ are the orbital angular and spin angular momenta,
respectively.  The simple step potential described in
Eq.~(\ref{scatter_V}) is presented in Fig.~(6). For parameters
typical of III-V semiconductors, the asymmetric and symmetric
scattering rates are almost flat as functions of the energy of the
incoming electron (see Figs.~ (4,5)). Comparing symmetric and
asymmetric scattering rates one can find directly from picture,
that for typical densities of 2DEG and typical Bohr radii in
semiconductors the ratio of symmetric scattering time to skew
scattering time is:
\begin{equation}\label{ratio_taus}
\frac{\tau }{\tau_{ss}}\approx 0.002\div 0.003
\end{equation}
We assumed, in above discussion, that the electron-impurity
potential is attractive, which is the most common case in
semiconductors, where ionized donors play the role of impurities.
However, the sign and the amplitude of skew scattering
contribution (and more precisely skew scattering rate) depends
strongly on the electron-impurity potential.

\subsection{Side jump current and resistivity matrix}
Now to determine the side jump contribution to the spin-Hall
effect we need to carefully define the spin-current. The
spin-current density operator must be calculated taking into
account the form of the physical velocity (see Eq.~(\ref{vphys2})):
\begin{eqnarray}\label{jspin}
\hat J_y^z &=&
-\frac{e}{2V}\sum_{i}\left(\upsilon^y_{phys,i}\sigma_{iz}+\sigma_{iz}\upsilon^y_{phys,i}\right)\nonumber\\
&=&\frac{-e}{V}\sum_i \left(\frac{p^y_{i}\sigma_{iz}}{m}+
\frac{2\alpha F_{ix}}{\hbar}\right)\nonumber\\ &\approx&
\frac{-e}{V} \sum_{i} \upsilon^y_{i}\sigma_{iz}
\end{eqnarray}
where the factor $2\alpha F_{ix}/\hbar$ vanishes  because the net
force $\vec F_i$ acting on an electron is zero when averaged over
a steady-state ensemble. Therefore, the spin component of the
current is
\begin{eqnarray}\label{jspin_sigma}
j_{\sigma} &=& -en_{\sigma}\vec{V_{\sigma}}\,,
\end{eqnarray}
 and using Eq.~(\ref{Vsigma}) for the drift velocity we obtain the
following form of side-jump current:
\begin{equation}\label{j_sj} \vec j^{sj} =
2e^2\alpha\sigma n_{\sigma}\vec E_{\sigma} \times z~,
\end{equation}
which evidently arises from the last term of Eq.~(\ref{Vsigma}).
This simple form of the side-jump current is valid only for
electrons in the conduction band. In this case, the side-jump
current depends only on the spin-orbit coupling, the density of
electrons and the spin-dependent electric field.  The complete
relation between the spin-component of current (from side jump and
skew scattering contributions) and electric field has the
following form:
\begin{equation}\label{electric_so}
\vec E_\sigma=\rho^D_{\sigma}\vec
j_{\sigma}+\sigma[\rho^{ss}_{\sigma}-
\lambda_{\sigma}\rho^D_{\sigma}]\vec j_{\sigma}\times \hat{z}
\end{equation}
where $\rho^D_{\sigma} =\frac{m}{n_{\sigma}e^2\tau_{\sigma}}$ is
the Drude resistivity, $\rho^{ss}_{\sigma}
=\frac{m}{n_{\sigma}e^2\tau^{ss}_{\sigma}}$ is the skew scattering
resistivity, and the last term in square brackets is the side-jump
contribution to the resistivity: $\lambda_{\sigma}
=\frac{2m\alpha}{\tau_{\sigma}}$. Eq.~(\ref{electric_so}) yields
to the following resistivity tensor (in the basis $x_{\uparrow}$,
$y_{\uparrow}$, $x_{\downarrow}$, $y_{\downarrow}$):
\begin{equation} \label{diagonal.resistivity} \rho=
\left(%
\begin{array}{llll}
     \rho^D_{\uparrow}
   & \rho^{ss}_{\uparrow}-
\lambda_{\uparrow}\rho^D_{\uparrow} & 0 & 0
\\-\rho^{ss}_{\uparrow}+
\lambda_{\uparrow}\rho^D_{\uparrow}& \rho^D_{\uparrow} & 0 & 0 \\
     0 & 0 & \rho^D_{\downarrow}
& -\rho^{ss}_{\downarrow}+ \lambda_{\downarrow}\rho^D_{\downarrow}
\\
     0 & 0 & \rho^{ss}_{\downarrow}-
\lambda_{\downarrow}\rho^D_{\downarrow} & \rho^D_{\downarrow}\\
\end{array}%
\right)
\end{equation}
The diagonal part of the resistivity reduces to the Drude formula
$\rho_{\sigma}^D =\frac{m}{ne^2\tau_{\sigma}}$, as expected.  The
spin-orbit interaction is entirely responsible for the appearance
of an off-diagonal (transverse) resistivity. The latter consists
of two competing terms associated with side-jump
($\lambda_{\sigma}\rho^D_{\sigma}$) and skew-scattering
($\rho^{ss}_{\sigma}$), as seen in
Eq.~(\ref{diagonal.resistivity}). The signs of side-jump and
skew-scattering terms are opposite for attractive
electron-impurity potential. Although, this is the most typical
case in doped semiconductors, it is important to emphasize that
side-jump and skew scattering terms  have equal signs in the
case of a repulsive electron-impurity potential. Also, as one can see from
tensor ~(\ref{diagonal.resistivity}) contributions scale
differently with the mobility. Since $\tau_{ss}\sim \tau$ the skew
scattering contribution to the resistivity is proportional to
$1/\mu$, where $\mu$ is a mobility, while the side jump
contribution scales as $1/\mu^2$. The opposite signs of two
contributions, and different scaling with mobility could allow to
distinguish between them in the experiments (see Section IV)
\cite{HankiewiczPRL06}. As expected, in the absence of
electron-electron interactions, the resistivity
tensor~(\ref{diagonal.resistivity}) does not include elements
between the opposite spins.

\section{Spin-Coulomb drag}
Ordinary Coulomb drag is caused by momentum exchange between electrons residing in two separate 2D layers and interacting via the Coulomb interaction (for review see \cite{Rojo99}).
The Spin-Coulomb drag is the single-layer analogue of the ordinary Coulomb drag.   In this case
spin-up and spin-down electrons play the role of electrons in different layers and
the friction arises (due to Coulomb
interactions) when spin-up and spin-down electrons move within one single layer
with different drift velocities \cite{Amico00}.

The simplest description of the spin-Coulomb drag is given in terms of a phenomenological
friction coefficient $\gamma$.  Later in the section we will
show that the Boltzmann equation approach confirms this phenomenological description.
Let us start with the equation of motion for the drift velocity of
spin-$\sigma$ electrons:
\begin{equation}\label{eqofmotion}
 m\dot{\vec V}_{\sigma} = -eN_{\sigma}\vec
 E_{\sigma}+\vec F_{\sigma,-\sigma}-\frac{m\vec
V_{\sigma}}{\tau_{\sigma}}+\frac{m\vec
V_{-\sigma}}{\tau'_{\sigma}}
\end{equation}
where $N_{\sigma}$ is the number of electrons with spin $\sigma$,
$\vec F_{\sigma,-\sigma}$ is the net force exerted by $-\sigma$
spins on $\sigma$ spins, $\frac{1}{\tau_{\sigma}}$
is the rate of change of momentum of electrons with spin $\sigma$
due to electron-impurity scattering and is basically the Drude
scattering rate, $\frac{\vec V_{-\sigma}}{\tau'_{\sigma}}$ is the
rate of change of momentum due to electron-impurity scattering in
which electron flip its spin from $-\sigma$ to $\sigma$. From
Newton's third law one immediately sees:
\begin{equation}\label{F}
\vec F_{\sigma,-\sigma} = - \vec F_{-\sigma,\sigma}
\end{equation}
and by Galilean  invariance this force  can only depend on the
relative  velocity  of the two components. Hence, for a weak
Coulomb coupling one writes:
\begin{equation}\label{Fcoulomb}
\vec F_{\sigma,-\sigma} = -\gamma N_{\sigma}(\vec
\upsilon_{i,\sigma}-\vec
\upsilon_{i,-\sigma})\frac{n_{-\sigma}}{n}
\end{equation}
where $n_{\sigma}$ is the density of electrons with spin $\sigma$
and $\gamma$ is a spin-drag scattering rate. Taking into account
Eq.~(\ref{jspin_sigma}) and applying Fourier transformation to
Eq.~(\ref{eqofmotion}) one gets following equation on the spin
current density:
\begin{eqnarray}\label{j_Fourier}
i\omega \vec j_{\sigma} (\omega) &=& \frac{-n_{\sigma} e^2 \vec
E_{\sigma}(\omega)}{m} + \left(\gamma\frac{n_{-\sigma}}{n}
+\frac{1}{\tau_{\sigma}}\right) \vec j_{\sigma}(\omega)
\nonumber\\ &-& \left(\gamma\frac{n_{\sigma}}{n}
+\frac{1}{\tau'_{\sigma}}\right) \vec j_{-\sigma}(\omega)
\end{eqnarray}
Inverting Eq~(\ref{j_Fourier})  gives us  electric field:
\begin{eqnarray}\label{E_Fourier}
\vec E_{\sigma}(\omega) &=&\left(\frac{-i\omega
m}{n_{\sigma}e^2}+\frac{m}{n_{\sigma}e^2\tau_{\sigma}}
+\frac{m\gamma}{ne^2}\frac{n_{-\sigma}}{n_{\sigma}}\right)\vec
j_{\sigma}(\omega) \nonumber\\
&-&\left(\frac{m}{n_{\sigma}e^2\tau'_{\sigma}}+\frac{m\gamma}{ne^2}\right)\vec
j_{-\sigma}(\omega)
\end{eqnarray}
From this we can immediately read the resistivity
tensor. Its real part, in the basis of $x_{\uparrow}$,
$x_{\downarrow}$,  has following form:
\begin{equation} \label{Coulomb.Rho.Matrix}\rho= \left(
\begin{array}{cc}
\rho^D_{\uparrow} + \rho^{SD}n_{\downarrow}/n_{\uparrow} &
 -\rho^{SD}-\rho^\prime_{\uparrow}
\\ -\rho^{SD}-\rho^\prime_{\downarrow} & \rho^D_{\downarrow} +
\rho^{SD}n_{\uparrow}/n_{\downarrow}\\
\end{array}%
\right)
\end{equation}
where $\rho^{SD} =m\gamma/ne^2$ is the spin Coulomb drag
resistivity and
$\rho^\prime_{\sigma}=m/n_{\sigma}e^2\tau^\prime_{\sigma}$.
Several features of this matrix are noteworthy. First the matrix
is symmetric. Second the off-diagonal terms are negative. The
minus sign can be easily explained. $\rho_{\uparrow\downarrow}$ is
the electric field induced in the up-spin channel by a current
flowing in the down-spin channel when the up spin current is zero.
Since a down spin current in the positive direction tends to drag
along the up-spins, a negative electric field is needed to
maintain the zero value of the up-spin current. There is no limit
on the magnitude of $\rho_{SD}$. The only restriction is  that the
eigenvalues of the real part of the resistivity matrix  should be
positive to ensure positivity of dissipation. Finally, the
spin-Coulomb drag appears in  both diagonal and off-diagonal terms so
the total contribution cancels to zero (in accordance with
Eq.~(\ref{F})) if the drift velocities of up and down spins are
equal.
\begin{figure}[thb]
\vskip 0.27 in
\includegraphics[width=3.4in]{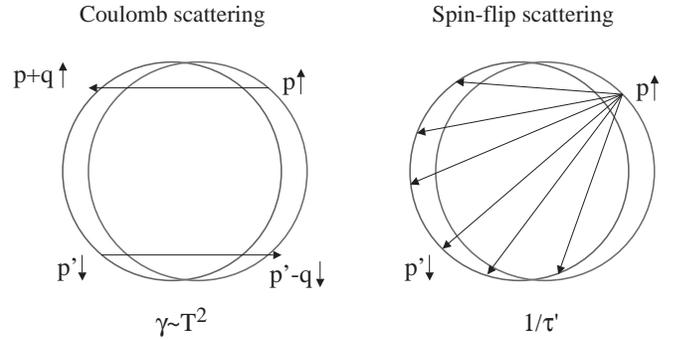}
\caption{Comparison of Coulomb scattering with the spin-flip
scattering. At finite temperature Coulomb scattering can be a more
effective mechanism of momentum exchange between up- and down-spin populations than spin-flip collisions with impurities.}
\end{figure}

Let us take a closer look at the competing off-diagonal terms:
spin-Coulomb drag and  spin-flip resistivities. At very low
temperature spin flip processes win  because in this limit the
Coulomb scattering is suppressed by phase space restrictions
(Pauli's exclusion principle) and $\gamma$ tends to zero as $T^2$
in three dimensions and $T^2lnT$ in 2D. However, the spin-flip
processes from electron-impurity collisions do not  effectively
contribute to momentum transfer between two spin-channels.  An
up-spin electron that collides with an impurity and flips its spin
orientation from up to down is almost equally likely to emerge in
any direction, as shown in Fig.~(7),  so the momentum transfer
from the up to the down spin orientation is minimal and
independent of what the down spins are doing.  However,   the
situation looks quite different for electron-electron collisions:
the collision of an up-spin electron with a  down-spin electron
leads to a momentum transfer that is preferentially oriented
against the relative velocity of the two electrons (see Fig.~(7))
and is proportional to the latter. Taking the spin-flip relaxation
time to be of  the order of 500ns (ten times larger than
spin-relaxation time in GaAs \cite{Kikkawa97,Kikkawa98}), and the
value of $\frac{1}{\gamma}$ of the order of the Drude scattering
time, around 1 ps \cite{Amico02,Amico03}, and temperatures  of the
order of the Fermi energy $T_F \sim 300K$ we estimate that the
spin-Coulomb drag contribution will dominate  already for $T \sim
0.3K$. Further, for mobilities typical for semiconductors 10$^4$
$\div$ 10$^5$ cm$^2$/Vs the ratio of spin-Coulomb drag to the
Drude resistivity can be as large as ten (see more detailed
discussion in the next Section and Fig.~(9)).

Returning to Boltzmann approach, the electron-electron
contribution to the collisional derivative has the
form\cite{Amico02}
\begin{widetext}
\begin{eqnarray}\label{coulomb.collision.integral}
\dot f_{\sigma}(\vec k)_{c,e-e}  \simeq -\sum_{\vec k' \vec p
\vec p'}W_C(\vec k \sigma,\vec p -\sigma;\vec k' \sigma, \vec p'
-\sigma) \left \{f_{\sigma}(\vec k)f_{-\sigma}(\vec
p)[1-f_{\sigma}(\vec k')] [1-f_{-\sigma}(\vec p' )] \right.
\nonumber\\ \left. - f_{\sigma}(\vec k')f_{-\sigma}(\vec
p')[1-f_{\sigma}(\vec k)] [1-f_{-\sigma}(\vec p)] \right\}
\delta_{\vec k +\vec p,\vec k'+\vec p'} \delta(\widetilde
\epsilon_{k\sigma}+\widetilde \epsilon_{p-\sigma}-\widetilde
\epsilon_{k'\sigma} - \widetilde \epsilon_{p'-\sigma})~,
\end{eqnarray}
\end{widetext}
where $W_C(\vec k \sigma,\vec p -\sigma;\vec k' \sigma, \vec p'
-\sigma)$ is the electron-electron scattering rate from $\vec k
\sigma, \vec p -\sigma$  to   $\vec k' \sigma, \vec p' -\sigma$,
and the Pauli factors $f_\sigma(\vec k)$,  $1-f_\sigma(\vec k')$
etc. ensure  that the initial states are occupied and the final
states empty as required by Pauli's exclusion principle. Notice
that, for our purposes, only collisions between electrons of
opposite spins are relevant, since collision between same-spin
electrons conserve the total momentum of each spin component.
After substituting the linearized Boltzmann equation into
Eq.~(\ref{coulomb.collision.integral}) and in the absence of
spin-orbit interactions one derives the following Coulomb collision
integral \cite{Hankiewicz05}:
\begin{widetext}
\begin{eqnarray}\label{coulomb.collision.integral3}
\dot f_{\sigma}(\vec k)_{c,e-e}  \simeq -\frac{1}{k_BT}\sum_{\vec
k' \vec p \vec p'}W_C(\vec k \sigma,\vec p -\sigma;\vec k' \sigma,
\vec p' -\sigma) [\hbar \vec V_{\sigma}-\hbar \vec
V_{-\sigma}](\vec k-\vec k')\\ \nonumber
 f_{0\sigma}(\epsilon_ k)f_{0-\sigma}(\epsilon_ p)f_{0
\sigma}(-\epsilon_{k'})f_{0-\sigma}(-\epsilon_{p'})
 \delta_{\vec k +\vec p,\vec k'+\vec p'} \delta(\epsilon_{k\sigma}+\epsilon_{p-\sigma}-\epsilon_{k'\sigma}
- \epsilon_{p'-\sigma})~,
\end{eqnarray}
\end{widetext}
where $T$ is the temperature, $k_B$ is the Boltzmann constant, and
we have made use of the identity $f_{0\sigma}(\epsilon_
k)f_{0-\sigma}(\epsilon_ p)[1-f_{0
\sigma}(\epsilon_{k'})][1-f_{0-\sigma}(\epsilon_{p'})]=[1-f_{0\sigma}(\epsilon_
k)][1-f_{0-\sigma}(\epsilon_ p)]f_{0
\sigma}(\epsilon_{k'})f_{0-\sigma}(\epsilon_{p'})$  for
$\epsilon_{k\sigma}+ \epsilon_{p-\sigma}- \epsilon_{k'\sigma} -
\epsilon_{p'-\sigma}=0$. The collision integral
Eq~(\ref{coulomb.collision.integral3}) is proportional to the
difference  of velocities for spin-up and spin-down electrons.
Therefore if a finite spin current is set up through the
application of an external field, then the Coulomb interaction
will tend to equalize the net momenta  of the two spin components,
causing $\langle V_{\uparrow}\rangle$ -$\langle
V_{\downarrow}\rangle$ to decay and thus can be interpreted, as we
explained before, as a damping mechanism for spin-current.

After substituting Eq.~(\ref{coulomb.collision.integral3}) into
self-consistent equation for a drift velocity
Eq.~(\ref{Boltzmann.Equation.2}), one obtains the spin-drag
coefficient $\gamma$ i.e. the rate of momentum transfer between up
and down spin electrons:
\begin{eqnarray} \label{gamma.Boltzmann}
\gamma &=& \frac{n}{n_\sigma n_{-\sigma}}\sum_{\vec k \vec k' \vec
p \vec p'}W_C(\vec k \sigma,\vec p -\sigma;\vec k' \sigma, \vec p'
-\sigma)\nonumber\\ &&\frac {(\vec k - \vec k')^2}{4 mk_BT}
f_{0\sigma}(\epsilon_ k)f_{0-\sigma}(\epsilon_ p)f_{0
\sigma}(-\epsilon_{k'})f_{0-\sigma}(-\epsilon_{p'})
\nonumber\\&\times& \delta_{\vec k +\vec p,\vec k'+\vec p'}
\delta(\epsilon_{k\sigma}+\epsilon_{p-\sigma}-\epsilon_{k'\sigma}
- \epsilon_{p'-\sigma}).
\end{eqnarray}
as well as equation of motion Eq.~(\ref{eqofmotion}) and
resistivity tensor Eq.~(\ref{Coulomb.Rho.Matrix}) derived before
through phenomenological approach.

Let us now describe briefly the idea of spin-grating experiments
\cite{Weber05}, where the spin-Coulomb drag has been observed.  A
periodic spin density can be induced by letting two linearly
polarized light beams coming from different directions interfere
on the surface of a two-dimensional electron gas. This
interference produces a spatially varying pattern of polarization,
with alternating regions of left-handed and right-handed circular
polarization separated by linearly polarized regions. The spin
density is optically induced in the regions of circular
polarization.  More precisely, the regions of right-handed
circular polarization have a larger up-spin density while the
regions of left-handed circular polarization have a larger
down-spin density. At a given time $t=0$ the pump light is turned
off and the subsequent time evolution of the spin-density is monitored by
Kerr spectroscopy. In Kerr spectroscopy, one measures the
amplitude of the spin-density modulation by looking at the
rotation of the plane of polarization of the light diffracted by
the spin-grating. The initial rate of decay of the spin grating
amplitude $\gamma_q$ depends on the wave vector q of the grating
in the following manner:\cite{Weber05}
\begin{eqnarray} \label{gamma_q}
\gamma_q =\frac{1}{\tau_s}+D_s q^2
\end{eqnarray}
where $\tau_s$ is the spin density relaxation time and $D_s$ is
the spin diffusion constant. Therefore $D_s$ can be found from
the slope of $\gamma_q$ vs. $q^2$. The spin diffusion constant in
the presence of spin-Coulomb drag was discussed in detail in
\cite{Amico01,Vignale2007} and has the following form:
\begin{eqnarray} \label{Ds}
\frac{D_s}{D^{0}_s} = \frac{\chi_s^{0}/\chi_s}{1+\gamma\tau}
\end{eqnarray}
where $D^0_{s}$ is the spin diffusion constant for the
non-interacting system, $\chi_s^0$ is the spin-susceptibility for
non-interacting system and $\chi_s$ is the interacting
spin-susceptibility, with Landau-Fermi-liquid corrections  taken into
account.

Actually, from the analysis presented in
Refs.~\cite{Amico01,Vignale2007} one expects that the
experimentally determined $D_s$ should include {\it two} effects:
the Fermi liquid correction to the spin
susceptibility~\cite{Thebook} and the spin-Coulomb drag
correction. However, the Fermi liquid correction to the spin
susceptibility is quite small.  It is given by the well-known
formula~\cite{Thebook}
\begin{equation}\label{Fermi_susc}
\frac{\chi^0_S}{\chi_S} = \frac{1+F_0^a}{\frac{m}{m_e}}
\end{equation}
where $m/m_e$ is the many-body mass enhancement and $F_0^a$ is the
Landau parameter described in detail in Ref.~\cite{Thebook}. Since
the $m/m_e \sim 0.96$ and $F_0^a \sim -0.2$, the interacting spin
susceptibility will be enhanced by no more than 20-30\% and
obviously will be independent of mobility of 2DEG. Therefore, the
Fermi liquid corrections to the spin-conductivity are very small
in comparison with the spin-Coulomb drag corrections and the
spin-Coulomb drag will be the main effect influencing the spin
transport.  Indeed, the experimentally determined
$D_s$~\cite{Weber05} was found to be in excellent agreement with
the theoretically predicted values for a strictly two-dimensional
electron gas in the random phase
approximation~\cite{Amico01,Amico03}. Following this, Badalyan
{\it et al.}~\cite{Badalyan08} noticed that the inclusion of the
finite thickness of the two-dimensional electron gas in the GaAs
quantum well  would worsen the agreement between theory and
experiment, because the form factor associated with the finite
thickness of the quantum well reduces the effective electron-electron
interaction at momentum transfers of the order of the Fermi
momentum, which are the most relevant for spin Coulomb drag.
Fortunately, it turned out that this reduction is compensated by
the inclusion of many-body effects beyond the random phase
approximation, namely local-field effects which, to a certain
extent, strengthen the effective Coulomb interaction by reducing
the electrostatic screening.\cite{Badalyan08} The final upshot of
the more careful analysis is that the theory remains in
quantitative agreement with experiment in a broad range of
temperatures.

\section{Influence of spin-Coulomb drag on the extrinsic spin-Hall
effect.}

\subsection{Resistance tensor}
In this Section we study the influence of electron-electron
interactions on the spin-Hall effect. Main discussion concerns
2DEG, however at the end of this Section we will comment on the
behavior of spin-Hall conductivity in bulk materials. We start
from the Hamiltonian which includes electron-electron
interactions:
\begin{equation}\label{Hint}
\hat H = \hat H_{ni} +\frac{1}{2} \sum_{i \neq
j}\frac{e^2}{\epsilon_b|\vec r_i-\vec r_j|}+\alpha \sum_i \vec p_i
\times \vec{\nabla}_i V^i_{ee}\cdot \vec \sigma_i
\end{equation}
where $\hat H_{ni}$ is defined by Eq.~(\ref{Hni}) and $V_{ee}
=\sum_{i \neq j}\frac{e^2}{\epsilon_b|\vec r_i-\vec r_j|}$. Notice
that the electric potential coming from electron-electron
interactions, like every potential whose gradient is non-zero,
generates the spin-orbit term in Hamiltonian.  This new spin-orbit
term, introduces  a new contribution to the Coulomb collision
integral: $2e\alpha\hbar\sigma (\vec E_{\sigma} + \vec
E_{-\sigma})\times \hat z (\vec k-\vec k')$ which adds up to
previous term i.e. the difference of velocities for spin-up and
down (see Eq.~(\ref{coulomb.collision.integral3})). As a
consequence, electron traveling say in x direction with spin up
can be scattered in a y direction with simultaneous spin-flip,
i.e. resistivity tensor contains terms which connect y and x
components with opposite spins. The full resistivity matrix in the
basis of $x_{\uparrow}$, $y_{\uparrow}$, $x_{\downarrow}$,
$y_{\downarrow}$ has the following form:
\begin{widetext}
\begin{equation} \label{Coulomb.Rho.Matrix}\rho= \left(
\begin{array}{llll}
\rho^D_{\uparrow} + \rho^{SD}n_{\downarrow}/n_{\uparrow} &
\rho^{ss}_{\uparrow}-
\lambda_{\uparrow}\rho^D_{\uparrow}+A^{\gamma\alpha}_{\uparrow} &
-\rho^{SD}-\rho^\prime_{\uparrow}& B^{\gamma\alpha}_{\uparrow}
\\-\rho^{ss}_{\uparrow}+
\lambda_{\uparrow}\rho^D_{\uparrow}-A^{\gamma\alpha}_{\uparrow}&
\rho^D_{\uparrow} +\rho^{SD}n_{\downarrow}/n_{\uparrow} &
-B^{\gamma\alpha}_{\uparrow} & -\rho^{SD}-\rho^\prime_{\uparrow}
\\ -\rho^{SD}-\rho^\prime_{\downarrow} & -B^{\gamma\alpha}_{\downarrow} & \rho^D_{\downarrow} +
\rho^{SD}n_{\uparrow}/n_{\downarrow}& -\rho^{ss}_{\downarrow}+
\lambda_{\downarrow}\rho^D_{\downarrow}-A^{\gamma\alpha}_{\downarrow}\\
 B^{\gamma\alpha}_{\downarrow}  & -\rho^{SD}-\rho^\prime_{\downarrow}  &
 \rho^{ss}_{\downarrow}-\lambda_{\downarrow}\rho^D_{\downarrow}+A^{\gamma\alpha}_{\downarrow}
&\rho^D_{\downarrow}+ \rho^{SD}n_{\uparrow}/n_{\downarrow}\\
\end{array}%
\right)
\end{equation}
\end{widetext}
where $\rho^{SD} =m\gamma/ne^2$ is the spin Coulomb drag
resistivity and
$\rho^\prime_{\sigma}=m/n_{\sigma}e^2\tau^\prime_{\sigma}$ (recall
that $\lambda_\sigma = \frac{2 m \alpha}{\tau_\sigma}$ is a
dimensionless quantity). $A^{\gamma\alpha}_{\sigma}$ and
$B^{\gamma\alpha}_{\sigma}$ represent the terms of the first order
in electron-electron coupling $\gamma$ and in SO coupling $\alpha$
and are defined as follows:
$A^{\gamma\alpha}_{\sigma}=-\lambda_{\sigma}\rho_{SD}n_{-\sigma}/n_{\sigma}
+2m\alpha\gamma[-n_{-\sigma}\rho^D_{\sigma}/n
+(n_{-\sigma}/n-n^2_{-\sigma}/nn_{\sigma})\rho^{SD}]$ and
$B^{\gamma\alpha}_{\sigma}=\lambda_{\sigma}\rho_{SD}+2m\alpha\gamma[-n_{-\sigma}\rho^D_{-\sigma}/n
+(n_{-\sigma}/n-n_{\sigma}/n)\rho^{SD}]$.
 Notice that the resistivity satisfies the following symmetry relations:
\begin{equation}\label{symmetry1}
\rho^{\beta \beta'}_{\sigma \sigma}= -\rho^{\beta' \beta}_{\sigma
\sigma}
\end{equation}
\begin{equation}\label{symmetry2}
\rho^{\beta \beta'}_{\sigma -{\sigma}}= \rho^{\beta'
\beta}_{-{\sigma} \sigma}
\end{equation}
where upper indices $\beta$ and $\beta'$ denote directions, and
the lower ones spin orientations. New features of the resistivity
matrix~\ref{Coulomb.Rho.Matrix} are the $\gamma\alpha$ and $\gamma^2
\alpha$ terms, which appear in the transverse elements of the resistivity
when the system is spin-polarized.  Furthermore, the off-diagonal resistivity elements
$\rho_{xy}^{\sigma -\sigma}$ are generally non zero. In the
paramagnetic case (zero spin polarization) the $\alpha\gamma^2$ terms are zero and the resistivity
matrix simplifies significantly. In this case, we find simple
interrelations between currents and electric fields in the spin and charge channels.
Omitting spin-flip processes ($1/\tau'=0$) we obtain
\begin{equation}\label{Ec}
\vec E_c = \rho^{D}\vec j_c+2(\rho^{ss}-\lambda
\rho^D-\lambda\rho_{SD})\vec j_s\times\hat{z}~,
\end{equation}
\begin{equation}\label{Es}
\vec E_s=4(\rho^{SD}+\rho^{D})\vec j_s+2(\rho^{ss}-\lambda
\rho^D-\lambda\rho_{SD})\vec j_c\times\hat{z}~,
\end{equation}
where the charge/spin components
of the electric field are defined as  $\vec{E}_c
=\frac{\vec{E}_{\uparrow}+\vec{E}_{\downarrow}}{2}$, $\vec{E}_s
=\vec{E}_{\uparrow}-\vec{E}_{\downarrow}$, and the charge and spin
currents are $\vec j_c =\vec j_{\uparrow}+\vec j_{\downarrow}$ and
$\vec j_{s} = \frac{\vec j_{\uparrow}-\vec j_{\downarrow}}{2}$,
respectively. The spin-Coulomb drag renormalizes the longitudinal
resistivity only in the spin channel. This is a consequence of the
fact that the net force exerted by spin-up electrons on spin-down
electrons is proportional to the difference of their drift
velocities, i.e. to the spin current. Additionally, the
electron-electron corrections  to the spin-orbit interactions
renormalize the transverse resistivity in the charge and spin
channels, so the Onsager relations between spin and charge
channels hold.
\begin{figure}[thb]
\vskip 0.27 in
\includegraphics[width=3.6in]{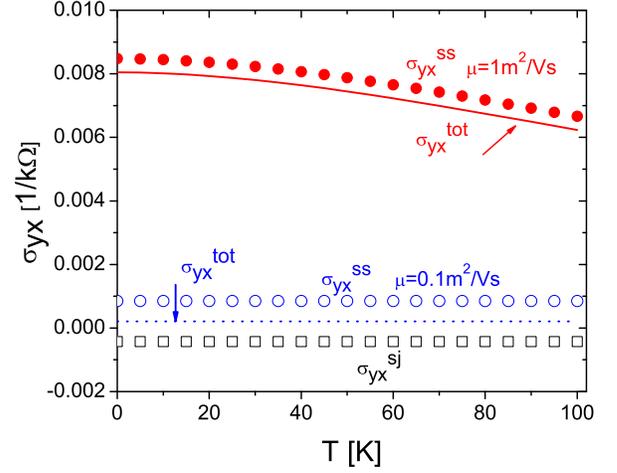}
\caption{Spin Hall conductivity as a function of temperature at
constant mobility. $\sigma_{yx}^{sj}$(open black squares),
$\sigma_{yx}^{ss}$ (open blue and close red
circles),$\sigma_{yx}^{tot}$ (red solid line and dashed blue line)
are the side-jump, skew-scattering and the total spin conductivity
in a presence of electron-electron interactions. Notice that only
skew scattering conductivity is modified by spin-Coulomb drag. Red
and blue curves/symbols are for $\mu =1$m$^2/$V.s, and $\mu
=0.1$m$^2/$V.s. We choose the parameters typical for the
experimental 2DEG confined in Al$_{0.1}$Ga$_{0.9}$As quantum well
i.e. density of electrons and impurities $n_{2D}=n_i=2.0\times
10^{12}$cm$^{-2}$, $m=0.074$m$_e$, and two sets of mobilities and
relaxation times: $\mu =0.1$m$^2/$V.s, $\tau =4\times 10^{-5}$ns
$\tau_{ss} = 0.02$ns and $\mu =1$m$^2/$V.s $\tau =4\times
10^{-4}$ns $\tau_{ss} = 0.2$ns. The effective spin-orbit coupling
$\alpha\hbar= 4.4{\AA}^2$ in accordance with \cite{Winkler2003}.
We used the model potential (see appendix) where an effective
impurity radius $a=9.45$nm, the height of attractive impurity
potential $V_0=-5$meV for $\mu =0.1$m$^2/$V.s and $V_0=-1.6$meV
for $\mu =1$m$^2/$V.s.}
\end{figure}
Under the assumption that the electric field is in the $x$
direction and has the same value for spin up and spin down
electrons we see that Eq.~(\ref{Ec}) and Eq.~(\ref{Es}) yield the
following formula for the spin current $j^z_{s,y}
=j_{\uparrow}-j_{\downarrow}$ in $y$ direction:
\begin{equation}\label{Jsy1}
j^{z}_{s,y} =
\left[\frac{\rho^{ss}/(\rho^D)^2}{1+\rho^{SD}/\rho^D}
-\frac{\lambda}{\rho_{D}}\right] E_x
\end{equation}
The first term in the square brackets  is associated with the
skew-scattering, while the second is the side-jump contribution.
Notice that the side-jump conductivity $\sigma^{sj} =
-\frac{\lambda}{\rho_{D}} = - 2\alpha n e^2$  depends neither on
the strength of disorder nor on the strength of the
electron-electron interaction. Moreover, as we showed
in~\cite{HankiewiczPRL06} by using a gauge invariance condition,
the side jump does not depend on the electron-impurity and
electron-electron scattering potential to all orders in both these
interactions. By contrast, the skew scattering contribution to the
spin conductivity, in the absence of e-e interactions scales with
transport scattering time. Therefore for very clean samples, the
skew scattering contribution would tend to infinity. However, this
unphysical behavior  is cured by the presence of spin-Coulomb
drag, which sets an upper limit to the spin conductivity of the
electron gas:  so the skew scattering term scales as
$\tau/(1+\rho_{SD}/\rho_{D}) = \tau/(1+\gamma\tau)$,  which tends
to a finite limit for $\tau \to \infty$. Let us now make an
estimate of skew scattering contribution to the spin-Hall
condutivity. Using the typical ratio of $\tau/\tau_{ss} \approx
10^{-3}$ (see Eq.~(\ref{ratio_taus}) we obtains:
\begin{equation}\label{sigma_ss2}
\sigma^{ss}= 10^{-3} \frac{\sigma_D}{1 + \gamma\tau}
\end{equation}
We will use this estimate in Section VI to compare importance of
different contributions to the spin-Hall conductivity.

Also, the direct dependence of the skew scattering conductivity
(see Eq.~(\ref{sigma_ss2})) on the transport scattering time
$\tau$ is the reason why this term is modified by spin-Coulomb
drag.  By contrast, the side-jump conductivity, which is
independent of $\tau$, remains completely unaffected. The total
spin Hall conductivity may either decrease or increase as a result
of the spin Coulomb drag, depending on the relative sign and size
of the skew-scattering and side-jump contributions. In the
ordinary case of attractive impurities, when the skew-scattering
contribution dominates, we expect an overall reduction in the
absolute value of the spin Hall conductivity.

This is shown in Fig.~(8). One can see that in high-mobility
samples the spin Coulomb drag  reduces the spin-Hall conductivity
very effectively. There is no upper limit to the  reduction of the
spin-Hall conductivity since the factor $1+\gamma\tau$ can become
arbitrarily large with increasing mobility. The behavior of
$\gamma\tau$ as a function of temperature is shown in Fig.~(9) for
a typical semiconductor mobility, $\mu=3\times 10^4$ cm$^2$/V.s.
For example, in a two-dimensional GaAs quantum well at a density
$n=10^{11}$ cm$^{-2}$ and mobility $3 \times 10^4$ cm$^2$/V.s the
factor $\gamma\tau$ is quite significantly larger than one in a
wide range of temperatures from $50$ K up to room temperature and
 above, and can substantially reduce the skew scattering term.

\begin{figure}[thb]
\vskip 0.27 in
\includegraphics[width=3.4in]{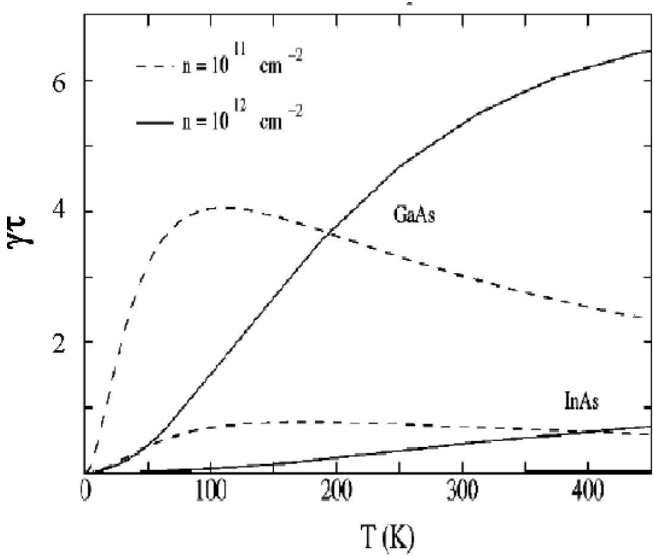}
\caption{The ratio of spin-Coulomb drag to Drude resistance
$\gamma\tau$ as a function of temperature for GaAs and InAs 2DEG
with mobility $\mu=3\times 10^4$ cm$^2$/V.s. The dashed and solid
lines correspond to $n=10^{11}cm^{-2}$ and $n=10^{12}cm^{-2}$,
respectively.}
\end{figure}

Let us finally comment on the spin-Hall conductivity in 3D.
Although the general formulas are the same as in 2D, the actual
value  of $\rho^{ss}$ must be obtained by solving a
three-dimensional scattering problem.  This has been done in
Ref.~\cite{Engel05} for the following model attractive potential
between the electron and an impurity atom:
\begin{equation}\label{3D}
V(r) =\frac{-e^{-q_sr}e^2}{\epsilon r}
\end{equation}
where $\epsilon$ is permittivity of material and $1/q_s$ is the
screening length associated with the Thomas-Fermi screening. For
this model potential, the spin-Hall conductivity takes the form:
\begin{equation}\label{sigmaSH_3D}
\sigma^{SH} = -2\alpha ne^2 + \frac{\gamma_s}{2} \sigma_D
\end{equation}
where $\sigma_D$ is the Drude conductivity and $\gamma_s$ is a
skewdness parameter, given approximately by $\gamma_s
=4\alpha\hbar/(a_B^2)$.\cite{Engel05} Notice that the side-jump
contribution  has exactly the same form as in 2D. Further, in 3D
the spin-Coulomb drag should modify the spin-Hall conductivity in
a similar way  i.e. by renormalizing the skew-scattering by factor
1+$\gamma\tau$ while leaving the side jump contribution to the
spin-conductivity unchanged. Therefore except for the different scaling
of $\gamma$ with temperature the spin-Hall conductivity behaves very
similarly in 2D and in 3D.\cite{Amico02,Amico03}

\subsection {Spin accumulation}
A quantitative theory of the spin accumulation in semiconductors
requires in general a proper treatment of the boundary conditions
as well as  electron-hole recombination effects \cite{Tse05}. In
this chapter we will study the influence of spin-Coulomb drag on
spin accumulation assuming that electrons are the only carriers
involved in transport. Our goal is to interpret the optical
experiments in which spin accumulation is measured
(\cite{Sih05,Kato04}). Notice that in previous theoretical papers
\cite{Engel05,Hankiewicz06}, directions of electric field and spin
accumulation where exchanged in relation to experimental ones
which led to a difference in a sign between experiment and
theoretical predictions due to the following relation between
resistivities: $\rho^{SH}_{xy} =-\rho^{SH}_{yx}$. In this review
we finally clarify this point and show that the sign of
experimental and theoretical spin-accumulations agree.

We consider a very long conductor in the form of a bar of length
$L$ in the $y$ direction and narrow width $W$ in the $x$
direction, exactly the same setup as in experiments
\cite{Sih05,Kato04} (see Fig~(10b)). A charge current flows only
in the $y$ direction. The spin components of the transverse
current $j_\sigma^x$, with $\sigma = \uparrow$ or $\downarrow$ add
up to zero everywhere and individually vanish on the edges of the
system, i.e. $ j_\sigma^x=0$ at $x = \pm W/2$. In order to satisfy
the boundary conditions the system cannot remain homogeneous in
the $x$-direction.  A position-dependent spin density, known as
{\it spin accumulation}  develops across the bar, and is reflected
in non-uniform chemical potentials $\mu_\sigma(x)$. In the steady
state regime the spatial derivative of the spin-current in the
$x$-direction must exactly balance the relaxation of the spin
density due to spin-flip processes i.e.:
\begin{equation}\label{spin_acc1}
\frac{e}{\sigma_s}\frac{dJ_{s}}{dx}
=\frac{\mu_{\sigma}(x)-\mu_{-\sigma}(x)}{L_s^2}
\end{equation}
where $L_s$ is the spin diffusion length and $\sigma_s$ is the
longitudinal spin-conductivity. Additionally, Ohm's law must be
fulfilled:
\begin{equation}\label{spin_acc2}
J_s =\sigma_s (E^x_{\sigma}-E^x_{-\sigma})
\end{equation}
where the effective electric field in the $x$- direction is
equivalent to the gradient of chemical potential:
\begin{equation}\label{spin_acc3}
eE^x_{\sigma}=d\mu_{\sigma}/dx~.
\end{equation}
Notice that in the limit of infinite spin-relaxation time ($L_s
\to \infty$) the divergence of spin-current equals zero and the
spin accumulation can be obtained directly from the homogeneous
formulas, Eqs.~(\ref{Ec}) and (\ref{Es}). In an inhomogenous case,
combined Eq.~(\ref{spin_acc1}) and Eq.~(\ref{spin_acc2}) lead to
the following equation for the spin accumulations \cite{Fert93}
\begin{equation}\label{spin_diffusion}
\frac{d^2[\mu_{\sigma}(x)-\mu_{-\sigma}(x)]}{d^2x}
=\frac{\mu_{\sigma}(x)-\mu_{-{\sigma}}(x)}{L_s^2}~,
\end{equation}
whose solution is:
\begin{equation}\label{spin_diffusion1a}
\mu_{\sigma}(x)-\mu_{-{\sigma}}(x)=Ce^{\frac{x}{L_s}}+
C'e^{-\frac{x}{L_s}}~,
\end{equation}
and C, C' are constants to be determined by the boundary
conditions $j^x_{\pm \sigma}(\pm W/2) =0$. Additionally using
Eq.~(\ref{spin_acc3}) and the resistivity tensor we can write the
boundary conditions for $E^x_{\sigma}(\pm W/2)$. Using the
boundary conditions for the spin-dependent chemical potentials and
the spin-dependent electric fields one finally finds the following
formula for the spin accumulation in a paramagnetic case:
\begin{eqnarray}\label{spin_diffusion4}
\mu_{\uparrow}(x)-\mu_{\downarrow}(x)= \frac{2eL_sE_y[\rho^{ss}-
\lambda\rho^D-\lambda\rho_{SD}]\sinh(x/L_s)}{\rho_D\cosh(W/2L_s)}~\nonumber\\.
\end{eqnarray}
The formula for the spin-accumulation in a spin-polarized case can
be also easily obtained and the interested reader can find it in
Ref.\cite{Hankiewicz06}. Finally, the spin-accumulation at the
edges of sample for $L=W/2$ has the form:
\begin{eqnarray}\label{spin_diffusion5}
&&V_{ac}^{x}=\mu_{\uparrow}(W/2L_s)-\mu_{\downarrow}(W/2L_s)=
\nonumber\\ && 2eL_sj_y[\rho^{ss}-
\lambda\rho^D-\lambda\rho_{SD}]\tanh(W/2L_s) ~.
\end{eqnarray}
The three terms in the square brackets of
Eq.~(\ref{spin_diffusion5}) are the skew-scattering term, the
ordinary side-jump contribution, and a Coulomb correction which
has its origin in the side-jump effect.  The latter is not a spin
Coulomb drag correction in the proper sense, for in this case the
transverse spin current, and hence the relative drift velocity of the electrons, is zero.
 What happens here is that the spin Hall current is canceled by an oppositely directed spin current,
 which is driven by the gradient of the spin chemical potential.
 Now the spin Hall current contains a universal contribution,
  the side-jump term, which is not affected by Coulomb interaction, but at the same time the constant of proportionality
  between the spin current and the gradient of the spin chemical potential, that is to say the longitudinal spin conductivity,
  is reduced by the Coulomb interaction.
Therefore, in order to maintain the balance against the unchanging
side-jump current, the absolute value of the gradient of the spin chemical potential
must increase when the Coulomb interaction is taken into account.
This effect may increase or decrease the total spin accumulation,
depending on the relative sign and magnitude of the side-jump and
skew scattering contributions. It reduces it in the common case, for an attractive electron-impurity potential,
where the two contributions have opposite signs and the
skew-scattering dominates.

\begin{figure}[thb]
\includegraphics[width=3.4in]{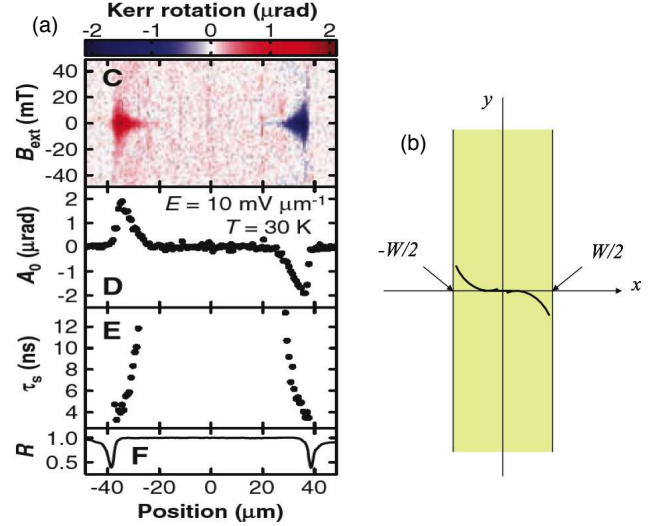}
\caption{(color online) (a) Spin accumulation as observed in
experiments by Kato et al. \cite{Kato04} (b) the spin accumulation
predicted theoretically (see Eq.~(\ref{spin_diffusion5})). For
experimental part of Figure: part (C) is a Kerr rotation as a
function of x and external magnetic field for electric field $E =
10 mV µm^{–1}$, parts (D and E) describe spatial dependence of
Kerr rotation peak $A_0$ and spin lifetime $\tau_s$ across the
channel, respectively. Part (F) shows the reflectivity R as a
function of x.}
\end{figure}
Additionally, Coulomb interactions affect the spin accumulation
indirectly through the spin diffusion length as shown in the equation below:
\begin{equation}\label{Ls}
L_s=\frac{\chi^0_s}{\chi_s}\frac{L_c}{1+\rho_{SD}/\rho_D}~,
\end{equation}
which follows immediately from Eq.~(\ref{Ds}). However, in the
limit of $W\ll L_s$, $\tanh(W/2L_s)$ can be approximated by
$W/2L_s$, and the spin accumulation at the edges becomes
independent of $L_s$. In this limit, the influence of the Coulomb
interaction on the spin accumulation is only through the
$\lambda\rho_{SD}$ term.

Let us now put in some numbers. For a two-dimensional electron gas
in an Al$_{0.1}$Ga$_{0.9}$As quantum well\cite{Sih05} with
electron and impurity concentrations $n_i=n_{2D} =2\times
10^{12}$cm$^{-2}$, mobility $\mu$=0.1m$^2$/V.s, $L_s=1\mu$m, $\tau
=4\times 10^{-5}$ns, $\tau_{ss} = 0.02$ns, $\alpha\hbar = 4.4
\AA^2$, j$_x$= 0.02 A/cm and for the sample with width
$W=100\mu$m, we calculate the spin accumulation to be
$-1.5$meV$/|e|$ on the right edge of the sample (relative to the
direction of the electric field) i.e. for $W/2= 50\mu m$. This
means that the non-equilibrium spin-density points down on the
right edge of the sample and up on the left edge exactly like in
the experiment.

The inhomogenous profile of spin-accumulation is presented in
Fig.~(10). Fig.~(10a) shows the signal of the spin-accumulation
(actually the Kerr rotation angle) observed in the experiment,
while Fig.~(10b) shows the profile of spin-accumulation expected
from the formula \ref{spin_diffusion5}. The general profile of the
spin accumulation is satisfactory on a qualitative level (taking
into account that we considered a very simple description of
spin-accumulation).

\begin{figure}[thb]
\includegraphics[width=3.4in]{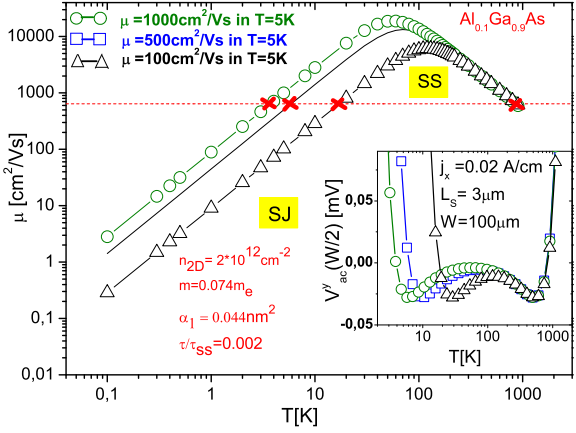}
\caption{(color online) Mobility, $\mu$, as a function of
temperature, $T$, for three different low-$T$ $\mu$'s. In inset,
the spin accumulation ($V^y_{ac}$) vs. $T$. The side jump
contribution to $V^y_{ac}$ dominates for low $T$. For increasing
$T$, the lower temperature red cross corresponds to the $T$ where
the sign of $V^y_{ac}$ starts to be controlled by skew scattering,
the higher temperature red cross to the place where side jump
dominates again.}
\end{figure}

As we mentioned before, it is possible to distinguish between side
jump and skew scattering contributions to the spin accumulation
because they scale differently with mobility. We have proposed an
experiment to distinguish between these two contributions in a
study of the temperature dependence of $\sigma^{SH}_{yx}$ or spin
accumulation $V^{x}_{ac}= -V^{y}_{ac}$, where the last equality stems from the fact
that the transverse resistivity elements are odd under exchange of spatial coordinates
i.e. $\rho^{xy}=-\rho^{yx}$. Fig.~(11) presents the behavior of
mobility versus temperature for experimentally attainable samples.
Due to different scattering mechanisms, the mobility scales non-monotonically with the temperature.
Hence $\mu$
will grow as $T^{3/2}$ for low $T$ as a result of scattering from
ionized impurities and will decrease as $T^{-3/2}$ for larger $T$
due to phonon scattering. It is thus possible to observe two
changes of sign of $V^{y}_{ac}$ moving from low to high $T$s. $\mu
=1/(AT^{-3/2}+BT^{3/2})$, where $A$ was found from the low-$T$
mobility and $B$ was fixed by a room temperature mobility of 0.3
m$^2$/Vs for AlGaAs. At low $T$ the mobility is low and the side
jump contribution to $V^{y}_{ac}$ dominates. With increasing $T$,
the first cross designates the point where the skew scattering
begins to dominate, and the second cross, at higher $T$, is the
point where the side jump takes control of the sign of
$V^{y}_{ac}$ again. Even if the sign change is not detected, by
measuring whether $V^{y}_{ac}$ increases or decreases as $\mu$
increases with changing $T$ it should possible to tell whether
side jump or skew scattering dominates.
Notice that the values of parameters for the theoretical curve
designated by circles are exactly the same as the values reported
for the samples in the recent experiments~\cite{Sih05} on a [110]
QW. The samples with lower mobilities can be easily obtained by
additional doping with Si inside the quantum well.

\section{Influence of Rashba type spin-orbit interaction on Spin-Coulomb drag}
So far, the influence of the Coulomb interaction on the intrinsic spin
Hall effect has not been analyzed. The issue is more complex than
the problem presented in Section IV i.e. the study of the
influence of spin-Coulomb drag on the extrinsic spin Hall effect.
The main difference between the problem with impurities and the
problem that considers spin-orbit interaction coming from the band
structure is that the latter usually does not conserve the
$z$-component of spin. The general form of intrinsic spin-orbit
interactions in 2D is:
\begin{eqnarray}\label{H_band}
  H_{b} = -\frac{1}{2}{\vec b} ({\vec k}) \cdot {\vec \sigma}
\end{eqnarray}
where $\vec \sigma$ is the vector of the Pauli matrices and $\vec
b(\vec k)$ is the intrinsic spin-orbit field. Due to time reversal
symmetry this field needs to fulfill the following condition: $\vec
b(\vec k)= -\vec b(-\vec k)$. For example for the simplest model
describing the spin-orbit interactions in a 2DEG oriented in [001]
direction one has\cite{Rashba84}:
\begin{eqnarray}\label{Hrashb}
 \vec b(\vec k) = 2\alpha \vec {\hat{z}}\times \vec k
\end{eqnarray}
where $\vec {\hat{z}}$ is the unit vector in z direction and
$\alpha$ is the spin-orbit coupling strength. This model is known
in literature as the Rashba model \cite{Rashba84}. For 2DHG, the
Luttinger Hamiltonian for low densities can be simplified by
taking into account only the heavy-hole band
\cite{Winkler00,Schliemann05}. This gives
\begin{eqnarray}\label{Hcubic}
 H_{2DHG}=i\alpha_h(k^3_{-}\sigma_{+} - k^3_{+}\sigma_{-})\,.
\end{eqnarray}
Obviously, these Hamiltonians do not conserve the $z$-component of
spin and therefore the simple approach presented in Section IV can
not be applied. The main complication  is that we need to consider
the whole density matrix (also the off-diagonal terms) in the
Boltzmann approach. However, the Rashba model (at least for the
standard definition of spin-current \cite{Niu06}) gives zero
spin-Hall conductivity
\cite{Inoue04,Shytov04,Dimitrova05,Raimondi05,Khaetskii06,Hankiewicz08} so we do not
expect that this result will be further modified by e-e
interactions. On the other hand, for the cubic spin-orbit models,
vertex corrections are not important and the spin-Hall
conductivity is of the order of $e^2/\hbar$. In this case, our
expectation is that the spin-Hall conductivity, being non-universal,  will be reduced by
spin-Coulomb drag. However, there has not been yet the calculations which
would quantitatively address this problem.

So far the only studied problem was the influence of spin-orbit
interactions on the spin-Coulomb drag \cite{Sarma07}. Diagrammatic
calculations \cite{Sarma07}, suggest that the spin-Coulomb drag is
actually enhanced by spin-orbit interaction of the Rashba type, at
least in the weak scattering regime. The correction is simply
additive to what we would expect from microscopic calculations of
spin-Coulomb drag without spin-orbit interaction \cite{Amico00}.
It has the form $ 3(\gamma_{int}^{*})^2$ where
$\gamma_{int}^{*}=\alpha m_e/(\upsilon_{F}m)$ where $\upsilon_{F}$
is the Fermi velocity. There is still  much room of course for
more detailed studies of the interplay between spin-orbit
interaction and spin Coulomb drag.  Recently, Weber {\it et
al.}~\cite{Weber07}  have undertaken experimental studies of the
relaxation of a spin grating in a two-dimensional electron gas
oriented in such a way that the spin-orbit interaction is
relevant.  A theoretical study by Weng {\it et al.},\cite{Weng08}
which takes into account both spin-orbit coupling and
electron-electron interaction concludes that the spin Coulomb drag
will still be visible as  a reduction of the spin diffusion
constant.  The latter is still determined from the initial decay
rate of the amplitude of the spin grating, but the full time
evolution of the amplitude involves two different spin relaxation
times for in-plane and out-of-plane dynamics respectively.

\section{Evolution of spin-Hall effect}
As we showed in previous Sections, the spin-Hall effect has
various contributions. Therefore, it is in place to compare the
importance of different mechanisms contributing to the spin-Hall
conductivity as a function of $\hbar\tau/m$. We chose
$\hbar\tau/m$ as a SHE evolution parameter  for two reasons (1) it
has a dimension of a squared length and can be directly compared
with the strength of the spin-orbit coupling $\alpha\hbar$ (2)
$\hbar\tau/m$ can be easily connected with the mobility. When
$\hbar\tau/m$ is expressed in $\AA^2$ it is approximately equal
$6\mu$, where $\mu$ is the mobility in cm$^2$/V.s.

In a d.c. limit we can distinguish three different regimes
\cite{Onoda06}: (1) ultraclean regime where $\frac{\hbar}{\tau}
\ll E_{so}$, where $E_{so}$ is the spin-orbit energy scale defined
as $E_{so} =E_F(\alpha\hbar/a_B^2)$, $a_B$ is the effective Bohr
radius (2) clean regime characterized by inequality $ E_{so} \ll
\frac{\hbar}{\tau} \ll E_F$, where $E_F$ is the Fermi energy, (3)
the dirty regime in which $\frac{\hbar}{\tau} > E_F$. In terms of
$\hbar\tau/m$ these three regimes correspond to (1)
$\frac{\hbar\tau}{m} \gg \frac{a_B^4}{\alpha\hbar}$ (ultraclean),
(2) $a_B^2 \ll \frac{\hbar\tau}{m} \ll \frac{a_B^4}{\alpha\hbar}$
(clean) and (3) $\frac{\hbar\tau}{m} <a_B^2$ (dirty) where we
assumed $n=a_B^{-3}$. On top of these limits we need to know the
relative importance of the skew scattering and side jump
contributions to the spin-Hall conductivity. Let us therefore
recall the final formulas for the skew scattering:
\begin{equation}\label{ssfinal}
\sigma^{ss}=10^{-3} \frac{\sigma_D}{1 + \gamma\tau}
\end{equation}
and side jump spin-Hall
conductivities:
\begin{equation}\label{sjfinal}
\sigma^{sj}= -2\alpha n e^2.
\end{equation}
The ratio of
these two conductivities is:
\begin{equation}\label{ratioSHE}
\frac{\sigma^{sj}}{\sigma^{ss}}\approx 3 \times 10^2
\frac{\alpha\hbar [\AA^2]}{\mu [cm^2/V.s]}
\end{equation}
where we used the connection between mobility  and $\hbar\tau/m$.
Also, in the above formula $\alpha\hbar$ is in units of \AA$^2$
while $\mu$ is in cm$^2$/V.s. The side jump and skew scattering
contributions have the same magnitude when: $\frac{\hbar\tau}{m}=
2\times 10^3 \alpha\hbar$. As we already mentioned the skew
scattering contribution scales with the mobility and therefore
will dominate the ultraclean regime, where this quantity is the
largest. The skew scattering contribution is cut-off by
spin-Coulomb drag when $\gamma \gg 1/\tau$ and the cut-off value
of spin-conductivity is $\sigma_{cut}^{ss} =
10^{-3}\frac{ne^2}{m\gamma}$. Intrinsic contribution (if it is not
zero, like in a Rashba model) dominates in a clean regime. As was
shown theoretically \cite{Zhang05}, the vertex corrections
connected with disorder are zero for p-doped semiconductors and
the spin-Hall conductivity is of the order of $e^2/\hbar$ and
therefore much larger than the side-jump contribution. For example
for typical 2D semiconducting hole gases with densities
(10$^{11}$cm$^{-2}$), side jump contribution is around thousand
times smaller than intrinsic one. Therefore in the scenario where
the skew-scattering regime passes to the side jump regime the
scale $2\times 10^3 \alpha\hbar $ must be larger than
$\frac{a_B^4}{\alpha \hbar}$. For doped semiconductors, for
example GaAs, the effective Bohr radius is 100\AA, so we have
$\frac{a_B^4}{\alpha \hbar}\approx 4\times 10^6 \alpha\hbar \gg
2\times 10^3 \alpha\hbar$. Therefore we will  have a direct
transition from the skew scattering to the intrinsic regime.
However, there are two options to observe the side-jump effect:
(i) the intrinsic contribution is zero (like in Rashba model),
then side-jump could be observed in the clean limit (ii) the
mobility is decreased by  changing the temperature and the
side-jump contribution could be observed in ultraclean regime as
we mentioned in Section V (see Fig.~(11)). In the dirty regime the
spin-Hall conductivity diminishes to zero. The scenario of
evolution of spin-Hall conductivity in semiconductors with
non-zero intrinsic contribution is presented in Fig.~(12).
In contrast, if one adopts Eq.~(\ref{ssfinal}) and Eq.~(\ref{sjfinal})
to describe the spin-Hall effect in metals, one finds
that the skew scattering term would evolve into
side-jump contribution for $\hbar\tau/m^* =10^3\alpha$ and the
intrinsic effect will eventually appear for $\hbar\tau/m^* =a_B^4/(\alpha\hbar)$ in the clean regime.
For parameters typical for Pt, the side jump and intrinsic contributions are of the same order and dominant.
Therefore further calculations (including the complexity of band structure) and experiments are needed to distinguish between various mechanisms contributing to SHE in metals.

\begin{figure}[thb]
\includegraphics[width=3.4in]{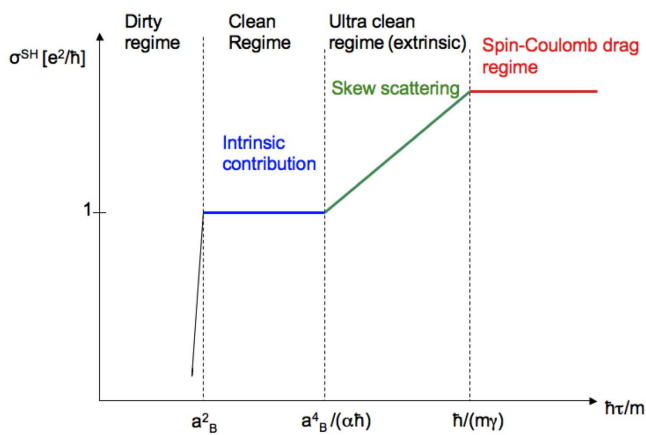}
\caption{(color online) Evolution of the spin-Hall effect as a
function of $\hbar\tau/m$ assuming that an electron-impurity scattering potential has the form of the screened Coulomb potential. The side jump could be observed in an
ultraclean regime if we decrease the mobility by for example the
tuning of temperature.}
\end{figure}

\section{Summary}
In this review we have summarized the current status of the
knowledge concerning the extrinsic spin Hall effect and the spin
Coulomb drag effect, and the relation between them. Careful
readers will notice that there are still plenty of open questions
and unsolved problems.

From the theoretical point of view, perhaps the most urgent open
challenge is the calculation of the influence of the spin-Coulomb
drag on the {\it intrinsic} spin Hall effect. From the
experimental point of view, it would be interesting to see
time-resolved studies of the spin Hall effect, possibly conducted
by spin-grating techniques~\cite{Weber05,Weber07} or by optical
spin injection techniques~\cite{Zhao06,Stern08}. Furthermore, a
direct detection of  the influence of the spin-Coulomb drag on the
spin-Hall effect and an experimental verification of the
theoretical predictions of  Section IV would be of great interest.
Finally, a full description of the interplay between spin-orbit
coupling and spin Coulomb drag remains an open challenge,
particularly at the experimental level.

{\it Acknowledgements.} This work was supported by NSF Grant No.
DMR-0705460.


\begin{thebibliography}{90}
\expandafter\ifx\csname natexlab\endcsname\relax\def\natexlab#1{#1}\fi
\expandafter\ifx\csname bibnamefont\endcsname\relax
  \def\bibnamefont#1{#1}\fi
\expandafter\ifx\csname bibfnamefont\endcsname\relax
  \def\bibfnamefont#1{#1}\fi
\expandafter\ifx\csname citenamefont\endcsname\relax
  \def\citenamefont#1{#1}\fi
\expandafter\ifx\csname url\endcsname\relax
  \def\url#1{\texttt{#1}}\fi
\expandafter\ifx\csname urlprefix\endcsname\relax\def\urlprefix{URL }\fi
\providecommand{\bibinfo}[2]{#2}
\providecommand{\eprint}[2][]{\url{#2}}

\bibitem[{\citenamefont{Awschalom and Flatte}(2007)}]{Awschalom07}
\bibinfo{author}{\bibfnamefont{D.~D.} \bibnamefont{Awschalom}}
  \bibnamefont{and} \bibinfo{author}{\bibfnamefont{M.~E.}
  \bibnamefont{Flatte}}, \bibinfo{journal}{Nature Physics}
  \textbf{\bibinfo{volume}{3}}, \bibinfo{pages}{153} (\bibinfo{year}{2007}).

\bibitem[{\citenamefont{Engel et~al.}(2007)\citenamefont{Engel, Rashba, and
  Halperin}}]{Handbook}
\bibinfo{author}{\bibfnamefont{H.-A.} \bibnamefont{Engel}},
  \bibinfo{author}{\bibfnamefont{E.~I.} \bibnamefont{Rashba}},
  \bibnamefont{and} \bibinfo{author}{\bibfnamefont{B.~I.}
  \bibnamefont{Halperin}}, \emph{\bibinfo{title}{Handbook of Magnetism and
  Advanced Magnetic Materials}} (\bibinfo{publisher}{John Wiley \& Sons Ltd,
  Chichester}, \bibinfo{address}{UK}, \bibinfo{year}{2007}).

\bibitem[{\citenamefont{Murakami}(2005)}]{Murakami05r}
\bibinfo{author}{\bibfnamefont{S.}~\bibnamefont{Murakami}},
  \bibinfo{journal}{Adv. in Solid State Phys.} \textbf{\bibinfo{volume}{45}},
  \bibinfo{pages}{197} (\bibinfo{year}{2005}).

\bibitem[{\citenamefont{Karplus and Luttinger}(1954)}]{Luttinger54}
\bibinfo{author}{\bibfnamefont{R.}~\bibnamefont{Karplus}} \bibnamefont{and}
  \bibinfo{author}{\bibfnamefont{J.~M.} \bibnamefont{Luttinger}},
  \bibinfo{journal}{Phys. Rev. B} \textbf{\bibinfo{volume}{95}},
  \bibinfo{pages}{1154} (\bibinfo{year}{1954}).

\bibitem[{\citenamefont{Smit}(1955)}]{Smit55}
\bibinfo{author}{\bibfnamefont{J.}~\bibnamefont{Smit}},
  \bibinfo{journal}{Physica} \textbf{\bibinfo{volume}{21}},
  \bibinfo{pages}{877} (\bibinfo{year}{1955}).

\bibitem[{\citenamefont{Smit}(1958)}]{Smit58}
\bibinfo{author}{\bibfnamefont{J.}~\bibnamefont{Smit}},
  \bibinfo{journal}{Physica} \textbf{\bibinfo{volume}{24}}, \bibinfo{pages}{39}
  (\bibinfo{year}{1958}).

\bibitem[{Ber({\natexlab{a}})}]{Berger70a}
\bibinfo{note}{L. Berger Phys. Rev. B {\bf 2}, 4559 (1970)}.

\bibitem[{Ber({\natexlab{b}})}]{Berger72}
\bibinfo{note}{L. Berger Phys. Rev. B {\bf 5}, 1862 (1972)}.

\bibitem[{\citenamefont{Lyo and Holstein}(1972)}]{Lyo72}
\bibinfo{author}{\bibfnamefont{S.~K.} \bibnamefont{Lyo}} \bibnamefont{and}
  \bibinfo{author}{\bibfnamefont{T.}~\bibnamefont{Holstein}},
  \bibinfo{journal}{Phys. Rev. Lett.} \textbf{\bibinfo{volume}{29}},
  \bibinfo{pages}{423} (\bibinfo{year}{1972}).

\bibitem[{\citenamefont{Nozi\'eres and Lewiner}(1973)}]{Nozieres}
\bibinfo{author}{\bibfnamefont{P.}~\bibnamefont{Nozi\'eres}} \bibnamefont{and}
  \bibinfo{author}{\bibfnamefont{C.}~\bibnamefont{Lewiner}},
  \bibinfo{journal}{J. Phys. (Paris)} \textbf{\bibinfo{volume}{34}},
  \bibinfo{pages}{901} (\bibinfo{year}{1973}).

\bibitem[{\citenamefont{Cr\'epieux and Bruno}(2001)}]{Bruno01}
\bibinfo{author}{\bibfnamefont{A.}~\bibnamefont{Cr\'epieux}} \bibnamefont{and}
  \bibinfo{author}{\bibfnamefont{P.}~\bibnamefont{Bruno}},
  \bibinfo{journal}{Phys. Rev. B} \textbf{\bibinfo{volume}{64}},
  \bibinfo{pages}{014416} (\bibinfo{year}{2001}).

\bibitem[{\citenamefont{Jungwirth et~al.}(2002)\citenamefont{Jungwirth, Niu,
  and MacDonald}}]{Jungwirth02}
\bibinfo{author}{\bibfnamefont{T.}~\bibnamefont{Jungwirth}},
  \bibinfo{author}{\bibfnamefont{Q.}~\bibnamefont{Niu}}, \bibnamefont{and}
  \bibinfo{author}{\bibfnamefont{A.~H.} \bibnamefont{MacDonald}},
  \bibinfo{journal}{Phys. Rev. Lett.} \textbf{\bibinfo{volume}{88}},
  \bibinfo{pages}{207208} (\bibinfo{year}{2002}).

\bibitem[{\citenamefont{Onoda and Nagaosa}(2003)}]{Onoda02}
\bibinfo{author}{\bibfnamefont{M.}~\bibnamefont{Onoda}} \bibnamefont{and}
  \bibinfo{author}{\bibfnamefont{N.}~\bibnamefont{Nagaosa}},
  \bibinfo{journal}{Phys. Rev. Lett.} \textbf{\bibinfo{volume}{90}},
  \bibinfo{pages}{206601} (\bibinfo{year}{2003}).

\bibitem[{\citenamefont{Dugaev et~al.}(2005)\citenamefont{Dugaev, Bruno,
  Taillefumier, Canals, and Lacroix}}]{Bruno05}
\bibinfo{author}{\bibfnamefont{V.~K.} \bibnamefont{Dugaev}},
  \bibinfo{author}{\bibfnamefont{P.}~\bibnamefont{Bruno}},
  \bibinfo{author}{\bibfnamefont{M.}~\bibnamefont{Taillefumier}},
  \bibinfo{author}{\bibfnamefont{B.}~\bibnamefont{Canals}}, \bibnamefont{and}
  \bibinfo{author}{\bibfnamefont{C.}~\bibnamefont{Lacroix}},
  \bibinfo{journal}{Phys. Rev. B} \textbf{\bibinfo{volume}{71}},
  \bibinfo{pages}{224423} (\bibinfo{year}{2005}).

\bibitem[{\citenamefont{Nunner et~al.}(2007)}]{Borunda07}
\bibinfo{author}{\bibfnamefont{T.~S.} \bibnamefont{Nunner}}
  \bibnamefont{et~al.}, \bibinfo{journal}{Phys. Rev. B}
  \textbf{\bibinfo{volume}{76}}, \bibinfo{pages}{235312}
  (\bibinfo{year}{2007}).

\bibitem[{\citenamefont{Sinitsyn}(2008)}]{Sinitsyn08}
\bibinfo{author}{\bibfnamefont{N.~A.} \bibnamefont{Sinitsyn}},
  \bibinfo{journal}{J. Phys. Cond. Matt.} \textbf{\bibinfo{volume}{20}},
  \bibinfo{pages}{023201} (\bibinfo{year}{2008}).

\bibitem[{\citenamefont{Dyakonov and Perel}(1971{\natexlab{a}})}]{Dyakonov71}
\bibinfo{author}{\bibfnamefont{M.~I.} \bibnamefont{Dyakonov}} \bibnamefont{and}
  \bibinfo{author}{\bibfnamefont{V.~I.} \bibnamefont{Perel}},
  \bibinfo{journal}{Phys. Lett. A} \textbf{\bibinfo{volume}{35}},
  \bibinfo{pages}{459} (\bibinfo{year}{1971}{\natexlab{a}}).

\bibitem[{\citenamefont{Dyakonov and Perel}(1971{\natexlab{b}})}]{Perel}
\bibinfo{author}{\bibfnamefont{M.~I.} \bibnamefont{Dyakonov}} \bibnamefont{and}
  \bibinfo{author}{\bibfnamefont{V.~I.} \bibnamefont{Perel}},
  \bibinfo{journal}{Zh. Eksp. Ter. Fiz.} \textbf{\bibinfo{volume}{13}},
  \bibinfo{pages}{657} (\bibinfo{year}{1971}{\natexlab{b}}).

\bibitem[{\citenamefont{Hirsch}(1999)}]{Hirsch99}
\bibinfo{author}{\bibfnamefont{J.~E.} \bibnamefont{Hirsch}},
  \bibinfo{journal}{Phys. Rev. Lett.} \textbf{\bibinfo{volume}{83}},
  \bibinfo{pages}{1834} (\bibinfo{year}{1999}).

\bibitem[{\citenamefont{Zhang}(2000)}]{Zhang00}
\bibinfo{author}{\bibfnamefont{S.}~\bibnamefont{Zhang}},
  \bibinfo{journal}{Phys. Rev. Lett.} \textbf{\bibinfo{volume}{85}},
  \bibinfo{pages}{393} (\bibinfo{year}{2000}).

\bibitem[{\citenamefont{Murakami et~al.}(2003)\citenamefont{Murakami, Nagaosa,
  and Zhang}}]{Murakami03}
\bibinfo{author}{\bibfnamefont{S.}~\bibnamefont{Murakami}},
  \bibinfo{author}{\bibfnamefont{N.}~\bibnamefont{Nagaosa}}, \bibnamefont{and}
  \bibinfo{author}{\bibfnamefont{S.-C.} \bibnamefont{Zhang}},
  \bibinfo{journal}{Science} \textbf{\bibinfo{volume}{301}},
  \bibinfo{pages}{1348} (\bibinfo{year}{2003}).

\bibitem[{\citenamefont{Sinova et~al.}(2004)\citenamefont{Sinova, Culcer, Niu,
  Sinitsyn, Jungwirth, and MacDonald}}]{Sinova04}
\bibinfo{author}{\bibfnamefont{J.}~\bibnamefont{Sinova}},
  \bibinfo{author}{\bibfnamefont{D.}~\bibnamefont{Culcer}},
  \bibinfo{author}{\bibfnamefont{Q.}~\bibnamefont{Niu}},
  \bibinfo{author}{\bibfnamefont{N.~A.} \bibnamefont{Sinitsyn}},
  \bibinfo{author}{\bibfnamefont{T.}~\bibnamefont{Jungwirth}},
  \bibnamefont{and} \bibinfo{author}{\bibfnamefont{A.~H.}
  \bibnamefont{MacDonald}}, \bibinfo{journal}{Phys. Rev. Lett.}
  \textbf{\bibinfo{volume}{92}}, \bibinfo{pages}{126603}
  (\bibinfo{year}{2004}).

\bibitem[{\citenamefont{Culcer et~al.}(2004)\citenamefont{Culcer, Sinova,
  Sinitsyn, Jungwirth, MacDonald, and Niu}}]{Culcer04}
\bibinfo{author}{\bibfnamefont{D.}~\bibnamefont{Culcer}},
  \bibinfo{author}{\bibfnamefont{J.}~\bibnamefont{Sinova}},
  \bibinfo{author}{\bibfnamefont{N.~A.} \bibnamefont{Sinitsyn}},
  \bibinfo{author}{\bibfnamefont{T.}~\bibnamefont{Jungwirth}},
  \bibinfo{author}{\bibfnamefont{A.~H.} \bibnamefont{MacDonald}},
  \bibnamefont{and} \bibinfo{author}{\bibfnamefont{Q.}~\bibnamefont{Niu}},
  \bibinfo{journal}{Phys. Rev. Lett.} \textbf{\bibinfo{volume}{93}},
  \bibinfo{pages}{046602} (\bibinfo{year}{2004}).

\bibitem[{\citenamefont{Schliemann and Loss}(2004)}]{Loss04}
\bibinfo{author}{\bibfnamefont{J.}~\bibnamefont{Schliemann}} \bibnamefont{and}
  \bibinfo{author}{\bibfnamefont{D.}~\bibnamefont{Loss}},
  \bibinfo{journal}{Phys. Rev. B} \textbf{\bibinfo{volume}{69}},
  \bibinfo{pages}{165315} (\bibinfo{year}{2004}).

\bibitem[{\citenamefont{Burkov et~al.}(2004)\citenamefont{Burkov, Nunez, and
  MacDonald}}]{Burkov04}
\bibinfo{author}{\bibfnamefont{A.~A.} \bibnamefont{Burkov}},
  \bibinfo{author}{\bibfnamefont{A.~S.} \bibnamefont{Nunez}}, \bibnamefont{and}
  \bibinfo{author}{\bibfnamefont{A.~H.} \bibnamefont{MacDonald}},
  \bibinfo{journal}{Phys. Rev. B} \textbf{\bibinfo{volume}{70}},
  \bibinfo{pages}{155308} (\bibinfo{year}{2004}).

\bibitem[{\citenamefont{Murakami et~al.}(2004)\citenamefont{Murakami, Nagaosa,
  and Zhang}}]{Murak04}
\bibinfo{author}{\bibfnamefont{S.}~\bibnamefont{Murakami}},
  \bibinfo{author}{\bibfnamefont{N.}~\bibnamefont{Nagaosa}}, \bibnamefont{and}
  \bibinfo{author}{\bibfnamefont{S.-C.} \bibnamefont{Zhang}},
  \bibinfo{journal}{Phys. Rev. B} \textbf{\bibinfo{volume}{69}},
  \bibinfo{pages}{235206} (\bibinfo{year}{2004}).

\bibitem[{\citenamefont{Murakami}(2004)}]{Murak041}
\bibinfo{author}{\bibfnamefont{S.}~\bibnamefont{Murakami}},
  \bibinfo{journal}{Phys. Rev. B} \textbf{\bibinfo{volume}{69}},
  \bibinfo{pages}{241202(R)} (\bibinfo{year}{2004}).

\bibitem[{\citenamefont{Sinitsyn et~al.}(2004)\citenamefont{Sinitsyn,
  Hankiewicz, Teizer, and Sinova}}]{Sinitsyn04}
\bibinfo{author}{\bibfnamefont{N.~A.} \bibnamefont{Sinitsyn}},
  \bibinfo{author}{\bibfnamefont{E.~M.} \bibnamefont{Hankiewicz}},
  \bibinfo{author}{\bibfnamefont{W.}~\bibnamefont{Teizer}}, \bibnamefont{and}
  \bibinfo{author}{\bibfnamefont{J.}~\bibnamefont{Sinova}},
  \bibinfo{journal}{Phys. Rev. B} \textbf{\bibinfo{volume}{70}},
  \bibinfo{pages}{081312(R)} (\bibinfo{year}{2004}).

\bibitem[{\citenamefont{Inoue et~al.}(2004)\citenamefont{Inoue, Bauer, and
  Molenkamp}}]{Inoue04}
\bibinfo{author}{\bibfnamefont{J.~I.} \bibnamefont{Inoue}},
  \bibinfo{author}{\bibfnamefont{G.~E.~W.} \bibnamefont{Bauer}},
  \bibnamefont{and} \bibinfo{author}{\bibfnamefont{L.~W.}
  \bibnamefont{Molenkamp}}, \bibinfo{journal}{Phys. Rev. B}
  \textbf{\bibinfo{volume}{70}}, \bibinfo{pages}{041303(R)}
  (\bibinfo{year}{2004}).

\bibitem[{\citenamefont{Mishchenko et~al.}(2004)\citenamefont{Mishchenko,
  Shytov, and Halperin}}]{Shytov04}
\bibinfo{author}{\bibfnamefont{E.~G.} \bibnamefont{Mishchenko}},
  \bibinfo{author}{\bibfnamefont{A.~V.} \bibnamefont{Shytov}},
  \bibnamefont{and} \bibinfo{author}{\bibfnamefont{B.~I.}
  \bibnamefont{Halperin}}, \bibinfo{journal}{Phys. Rev. Lett.}
  \textbf{\bibinfo{volume}{93}}, \bibinfo{pages}{226602}
  (\bibinfo{year}{2004}).

\bibitem[{\citenamefont{Schliemann and Loss}(2005)}]{Schliemann05}
\bibinfo{author}{\bibfnamefont{J.}~\bibnamefont{Schliemann}} \bibnamefont{and}
  \bibinfo{author}{\bibfnamefont{D.}~\bibnamefont{Loss}},
  \bibinfo{journal}{Phys. Rev. B} \textbf{\bibinfo{volume}{71}},
  \bibinfo{pages}{085308} (\bibinfo{year}{2005}).

\bibitem[{\citenamefont{Dimitrova}(2005)}]{Dimitrova05}
\bibinfo{author}{\bibfnamefont{O.~V.} \bibnamefont{Dimitrova}},
  \bibinfo{journal}{Phys. Rev. B} \textbf{\bibinfo{volume}{71}},
  \bibinfo{pages}{245327} (\bibinfo{year}{2005}).

\bibitem[{\citenamefont{Raimondi and Schwab}(2005)}]{Raimondi05}
\bibinfo{author}{\bibfnamefont{R.}~\bibnamefont{Raimondi}} \bibnamefont{and}
  \bibinfo{author}{\bibfnamefont{P.}~\bibnamefont{Schwab}},
  \bibinfo{journal}{Phys. Rev. B} \textbf{\bibinfo{volume}{71}},
  \bibinfo{pages}{033311} (\bibinfo{year}{2005}).

\bibitem[{\citenamefont{Khaetskii}(2006)}]{Khaetskii06}
\bibinfo{author}{\bibfnamefont{A.}~\bibnamefont{Khaetskii}},
  \bibinfo{journal}{Phys. Rev. B} \textbf{\bibinfo{volume}{73}},
  \bibinfo{pages}{115323} (\bibinfo{year}{2006}).

\bibitem[{\citenamefont{Duckheim and Loss}(2006)}]{Loss06}
\bibinfo{author}{\bibfnamefont{M.}~\bibnamefont{Duckheim}} \bibnamefont{and}
  \bibinfo{author}{\bibfnamefont{D.}~\bibnamefont{Loss}},
  \bibinfo{journal}{Nature Physics} \textbf{\bibinfo{volume}{2}},
  \bibinfo{pages}{195} (\bibinfo{year}{2006}).

\bibitem[{\citenamefont{Gorini et~al.}(2008)\citenamefont{Gorini, Schwab,
  Dzierzawa, and Raimondi}}]{Gorini08}
\bibinfo{author}{\bibfnamefont{C.}~\bibnamefont{Gorini}},
  \bibinfo{author}{\bibfnamefont{P.}~\bibnamefont{Schwab}},
  \bibinfo{author}{\bibfnamefont{M.}~\bibnamefont{Dzierzawa}},
  \bibnamefont{and} \bibinfo{author}{\bibfnamefont{R.}~\bibnamefont{Raimondi}},
  \bibinfo{journal}{Phys. Rev. B} \textbf{\bibinfo{volume}{78}},
  \bibinfo{pages}{125327} (\bibinfo{year}{2008}).

\bibitem[{\citenamefont{Tse and Sarma}(2006)}]{Sarm06}
\bibinfo{author}{\bibfnamefont{W.~K.} \bibnamefont{Tse}} \bibnamefont{and}
  \bibinfo{author}{\bibfnamefont{S.~D.} \bibnamefont{Sarma}},
  \bibinfo{journal}{Phys. Rev. B} \textbf{\bibinfo{volume}{74}},
  \bibinfo{pages}{245309} (\bibinfo{year}{2006}).

\bibitem[{\citenamefont{Hankiewicz and Vignale}(2008)}]{Hankiewicz08}
\bibinfo{author}{\bibfnamefont{E.~M.} \bibnamefont{Hankiewicz}}
  \bibnamefont{and} \bibinfo{author}{\bibfnamefont{G.}~\bibnamefont{Vignale}},
  \bibinfo{journal}{Phys. Rev. Lett.} \textbf{\bibinfo{volume}{100}},
  \bibinfo{pages}{026602} (\bibinfo{year}{2008}).

\bibitem[{\citenamefont{Kato et~al.}(2004)\citenamefont{Kato, Myers, Gossard,
  and Awschalom}}]{Kato04}
\bibinfo{author}{\bibfnamefont{Y.~K.} \bibnamefont{Kato}},
  \bibinfo{author}{\bibfnamefont{R.~C.} \bibnamefont{Myers}},
  \bibinfo{author}{\bibfnamefont{A.~C.} \bibnamefont{Gossard}},
  \bibnamefont{and} \bibinfo{author}{\bibfnamefont{D.~D.}
  \bibnamefont{Awschalom}}, \bibinfo{journal}{Science}
  \textbf{\bibinfo{volume}{306}}, \bibinfo{pages}{1910} (\bibinfo{year}{2004}).

\bibitem[{\citenamefont{Sih et~al.}(2005)}]{Sih05}
\bibinfo{author}{\bibfnamefont{V.}~\bibnamefont{Sih}} \bibnamefont{et~al.},
  \bibinfo{journal}{Nature Physics} \textbf{\bibinfo{volume}{1}},
  \bibinfo{pages}{31} (\bibinfo{year}{2005}).

\bibitem[{\citenamefont{Wunderlich et~al.}(2005)}]{Wunderlich05}
\bibinfo{author}{\bibfnamefont{J.}~\bibnamefont{Wunderlich}}
  \bibnamefont{et~al.}, \bibinfo{journal}{Phys. Rev. Lett.}
  \textbf{\bibinfo{volume}{94}}, \bibinfo{pages}{047204}
  (\bibinfo{year}{2005}).

\bibitem[{\citenamefont{Stern et~al.}(2006)}]{Stern06}
\bibinfo{author}{\bibfnamefont{N.~P.} \bibnamefont{Stern}}
  \bibnamefont{et~al.}, \bibinfo{journal}{Phys. Rev. Lett.}
  \textbf{\bibinfo{volume}{97}}, \bibinfo{pages}{126603}
  (\bibinfo{year}{2006}).

\bibitem[{\citenamefont{Stern et~al.}(2008)\citenamefont{Stern, Steuerman,
  Mack, Gossard, and Awschalom}}]{Stern08}
\bibinfo{author}{\bibfnamefont{N.~P.} \bibnamefont{Stern}},
  \bibinfo{author}{\bibfnamefont{D.~W.} \bibnamefont{Steuerman}},
  \bibinfo{author}{\bibfnamefont{S.}~\bibnamefont{Mack}},
  \bibinfo{author}{\bibfnamefont{A.~C.} \bibnamefont{Gossard}},
  \bibnamefont{and} \bibinfo{author}{\bibfnamefont{D.~D.}
  \bibnamefont{Awschalom}}, \bibinfo{journal}{Nature Physics}
  \textbf{\bibinfo{volume}{4}}, \bibinfo{pages}{843} (\bibinfo{year}{2008}).

\bibitem[{\citenamefont{Zhao et~al.}(2006)}]{Zhao06}
\bibinfo{author}{\bibfnamefont{H.}~\bibnamefont{Zhao}} \bibnamefont{et~al.},
  \bibinfo{journal}{Phys. Rev. Lett.} \textbf{\bibinfo{volume}{96}},
  \bibinfo{pages}{246601} (\bibinfo{year}{2006}).

\bibitem[{\citenamefont{Hankiewicz et~al.}(2004)\citenamefont{Hankiewicz,
  Molenkamp, Jungwirth, and Sinova}}]{Hankiewicz04}
\bibinfo{author}{\bibfnamefont{E.~M.} \bibnamefont{Hankiewicz}},
  \bibinfo{author}{\bibfnamefont{L.~W.} \bibnamefont{Molenkamp}},
  \bibinfo{author}{\bibfnamefont{T.}~\bibnamefont{Jungwirth}},
  \bibnamefont{and} \bibinfo{author}{\bibfnamefont{J.}~\bibnamefont{Sinova}},
  \bibinfo{journal}{Phys. Rev. B} \textbf{\bibinfo{volume}{70}},
  \bibinfo{pages}{241301(R)} (\bibinfo{year}{2004}).

\bibitem[{\citenamefont{Br\"{u}ne et~al.}(2008)\citenamefont{Br\"{u}ne, Roth,
  Novik, K\"onig, Buhmann, Hankiewicz, Hanke, Sinova, and Molenkamp}}]{Brune08}
\bibinfo{author}{\bibfnamefont{C.}~\bibnamefont{Br\"{u}ne}},
  \bibinfo{author}{\bibfnamefont{A.}~\bibnamefont{Roth}},
  \bibinfo{author}{\bibfnamefont{E.~G.} \bibnamefont{Novik}},
  \bibinfo{author}{\bibfnamefont{M.}~\bibnamefont{K\"onig}},
  \bibinfo{author}{\bibfnamefont{H.}~\bibnamefont{Buhmann}},
  \bibinfo{author}{\bibfnamefont{E.~M.} \bibnamefont{Hankiewicz}},
  \bibinfo{author}{\bibfnamefont{W.}~\bibnamefont{Hanke}},
  \bibinfo{author}{\bibfnamefont{J.}~\bibnamefont{Sinova}}, \bibnamefont{and}
  \bibinfo{author}{\bibfnamefont{L.~W.} \bibnamefont{Molenkamp}}
  (\bibinfo{year}{2008}), \eprint{arXiv:0812.3768}.

\bibitem[{\citenamefont{Gui et~al.}(2004)}]{Gui04}
\bibinfo{author}{\bibfnamefont{Y.~S.} \bibnamefont{Gui}} \bibnamefont{et~al.},
  \bibinfo{journal}{Phys. Rev. B} \textbf{\bibinfo{volume}{70}},
  \bibinfo{pages}{115328} (\bibinfo{year}{2004}).

\bibitem[{\citenamefont{Bychkov and Rashba}(1984)}]{Rashba84}
\bibinfo{author}{\bibfnamefont{Y.~A.} \bibnamefont{Bychkov}} \bibnamefont{and}
  \bibinfo{author}{\bibfnamefont{E.~I.} \bibnamefont{Rashba}},
  \bibinfo{journal}{J. Phys. C} \textbf{\bibinfo{volume}{17}},
  \bibinfo{pages}{6039} (\bibinfo{year}{1984}).

\bibitem[{\citenamefont{Hankiewicz et~al.}(2005)\citenamefont{Hankiewicz, Li,
  Jungwirth, Niu, Shen, and Sinova}}]{Hankiewicz05}
\bibinfo{author}{\bibfnamefont{E.~M.} \bibnamefont{Hankiewicz}},
  \bibinfo{author}{\bibfnamefont{J.}~\bibnamefont{Li}},
  \bibinfo{author}{\bibfnamefont{T.}~\bibnamefont{Jungwirth}},
  \bibinfo{author}{\bibfnamefont{Q.}~\bibnamefont{Niu}},
  \bibinfo{author}{\bibfnamefont{S.-Q.} \bibnamefont{Shen}}, \bibnamefont{and}
  \bibinfo{author}{\bibfnamefont{J.}~\bibnamefont{Sinova}},
  \bibinfo{journal}{Phys. Rev. B} \textbf{\bibinfo{volume}{72}},
  \bibinfo{pages}{155305} (\bibinfo{year}{2005}).

\bibitem[{\citenamefont{Valenzuela and Tinkham}(2006)}]{Tinkham06}
\bibinfo{author}{\bibfnamefont{S.~O.} \bibnamefont{Valenzuela}}
  \bibnamefont{and} \bibinfo{author}{\bibfnamefont{M.}~\bibnamefont{Tinkham}},
  \bibinfo{journal}{Nature} \textbf{\bibinfo{volume}{442}},
  \bibinfo{pages}{176} (\bibinfo{year}{2006}).

\bibitem[{\citenamefont{Weng et~al.}()\citenamefont{Weng, Chandrasekhar,
  Miniatura, and Englert}}]{Englert08}
\bibinfo{author}{\bibfnamefont{K.~C.} \bibnamefont{Weng}},
  \bibinfo{author}{\bibfnamefont{N.}~\bibnamefont{Chandrasekhar}},
  \bibinfo{author}{\bibfnamefont{C.}~\bibnamefont{Miniatura}},
  \bibnamefont{and} \bibinfo{author}{\bibfnamefont{B.-G.}
  \bibnamefont{Englert}}, \eprint{arXiv:0804.0096}.

\bibitem[{\citenamefont{Shchelushkin and Brataas}(2005)}]{Brataas05}
\bibinfo{author}{\bibfnamefont{R.~V.} \bibnamefont{Shchelushkin}}
  \bibnamefont{and} \bibinfo{author}{\bibfnamefont{A.}~\bibnamefont{Brataas}},
  \bibinfo{journal}{Phys. Rev. B} \textbf{\bibinfo{volume}{72}},
  \bibinfo{pages}{073110} (\bibinfo{year}{2005}).

\bibitem[{\citenamefont{Tanaka et~al.}(2008)\citenamefont{Tanaka, Kontani,
  Naito, Naito, Hirashima, K.Yamada, and Inoue}}]{Tanaka08}
\bibinfo{author}{\bibfnamefont{T.}~\bibnamefont{Tanaka}},
  \bibinfo{author}{\bibfnamefont{H.}~\bibnamefont{Kontani}},
  \bibinfo{author}{\bibfnamefont{M.}~\bibnamefont{Naito}},
  \bibinfo{author}{\bibfnamefont{T.}~\bibnamefont{Naito}},
  \bibinfo{author}{\bibfnamefont{D.~S.} \bibnamefont{Hirashima}},
  \bibinfo{author}{\bibnamefont{K.Yamada}}, \bibnamefont{and}
  \bibinfo{author}{\bibfnamefont{J.}~\bibnamefont{Inoue}},
  \bibinfo{journal}{Phys. Rev. B} \textbf{\bibinfo{volume}{77}},
  \bibinfo{pages}{165117} (\bibinfo{year}{2008}).

\bibitem[{\citenamefont{Guo et~al.}()\citenamefont{Guo, Murakami, Chen, and
  Nagaosa}}]{Murakami08}
\bibinfo{author}{\bibfnamefont{G.~Y.} \bibnamefont{Guo}},
  \bibinfo{author}{\bibfnamefont{S.}~\bibnamefont{Murakami}},
  \bibinfo{author}{\bibfnamefont{T.-W.} \bibnamefont{Chen}}, \bibnamefont{and}
  \bibinfo{author}{\bibfnamefont{N.}~\bibnamefont{Nagaosa}},
  \eprint{arXiv:0705.0409v4}.

\bibitem[{\citenamefont{Kane and Mele}(2005)}]{Kane05}
\bibinfo{author}{\bibfnamefont{C.~L.} \bibnamefont{Kane}} \bibnamefont{and}
  \bibinfo{author}{\bibfnamefont{E.~J.} \bibnamefont{Mele}},
  \bibinfo{journal}{Phys. Rev. Lett.} \textbf{\bibinfo{volume}{95}},
  \bibinfo{pages}{226801} (\bibinfo{year}{2005}).

\bibitem[{\citenamefont{Bernevig and Zhang}(2006)}]{ZhangQSHE06}
\bibinfo{author}{\bibfnamefont{B.~A.} \bibnamefont{Bernevig}} \bibnamefont{and}
  \bibinfo{author}{\bibfnamefont{S.-C.} \bibnamefont{Zhang}},
  \bibinfo{journal}{Phys. Rev. Lett.} \textbf{\bibinfo{volume}{96}},
  \bibinfo{pages}{106802} (\bibinfo{year}{2006}).

\bibitem[{\citenamefont{Bernevig et~al.}(2006)\citenamefont{Bernevig, Hughes,
  and Zhang}}]{ZhangSC06}
\bibinfo{author}{\bibfnamefont{B.~A.} \bibnamefont{Bernevig}},
  \bibinfo{author}{\bibfnamefont{T.~L.} \bibnamefont{Hughes}},
  \bibnamefont{and} \bibinfo{author}{\bibfnamefont{S.-C.} \bibnamefont{Zhang}},
  \bibinfo{journal}{Science} \textbf{\bibinfo{volume}{314}},
  \bibinfo{pages}{1757} (\bibinfo{year}{2006}).

\bibitem[{\citenamefont{K\"{o}nig et~al.}(2007)\citenamefont{K\"{o}nig,
  Wiedmann, Br\"{u}ne, Roth, Buhmann, Molenkamp, Qi, and Zhang}}]{Konig07}
\bibinfo{author}{\bibfnamefont{M.}~\bibnamefont{K\"{o}nig}},
  \bibinfo{author}{\bibfnamefont{S.}~\bibnamefont{Wiedmann}},
  \bibinfo{author}{\bibfnamefont{C.}~\bibnamefont{Br\"{u}ne}},
  \bibinfo{author}{\bibfnamefont{A.}~\bibnamefont{Roth}},
  \bibinfo{author}{\bibfnamefont{H.}~\bibnamefont{Buhmann}},
  \bibinfo{author}{\bibfnamefont{L.~W.} \bibnamefont{Molenkamp}},
  \bibinfo{author}{\bibfnamefont{X.~L.} \bibnamefont{Qi}}, \bibnamefont{and}
  \bibinfo{author}{\bibfnamefont{S.-C.} \bibnamefont{Zhang}},
  \bibinfo{journal}{Science} \textbf{\bibinfo{volume}{318}},
  \bibinfo{pages}{766} (\bibinfo{year}{2007}).

\bibitem[{\citenamefont{K\"{o}nig et~al.}(2008)\citenamefont{K\"{o}nig,
  Buhmann, Molenkamp, Hughes, Liu, Qi, and Zhang}}]{Konig08}
\bibinfo{author}{\bibfnamefont{M.}~\bibnamefont{K\"{o}nig}},
  \bibinfo{author}{\bibfnamefont{H.}~\bibnamefont{Buhmann}},
  \bibinfo{author}{\bibfnamefont{L.~W.} \bibnamefont{Molenkamp}},
  \bibinfo{author}{\bibfnamefont{T.~L.} \bibnamefont{Hughes}},
  \bibinfo{author}{\bibfnamefont{C.~X.} \bibnamefont{Liu}},
  \bibinfo{author}{\bibfnamefont{X.~L.} \bibnamefont{Qi}}, \bibnamefont{and}
  \bibinfo{author}{\bibfnamefont{S.-C.} \bibnamefont{Zhang}},
  \bibinfo{journal}{J. of Phys. Soc. of Japan} \textbf{\bibinfo{volume}{77}},
  \bibinfo{pages}{031007} (\bibinfo{year}{2008}).

\bibitem[{\citenamefont{Wu et~al.}(2006)\citenamefont{Wu, Bernevig, and
  Zhang}}]{Wu06}
\bibinfo{author}{\bibfnamefont{C.}~\bibnamefont{Wu}},
  \bibinfo{author}{\bibfnamefont{B.~A.} \bibnamefont{Bernevig}},
  \bibnamefont{and} \bibinfo{author}{\bibfnamefont{S.-C.} \bibnamefont{Zhang}},
  \bibinfo{journal}{Phys. Rev. Lett.} \textbf{\bibinfo{volume}{96}},
  \bibinfo{pages}{106401} (\bibinfo{year}{2006}).

\bibitem[{\citenamefont{Xu and Moore}(2006)}]{Xu06}
\bibinfo{author}{\bibfnamefont{C.}~\bibnamefont{Xu}} \bibnamefont{and}
  \bibinfo{author}{\bibfnamefont{J.}~\bibnamefont{Moore}},
  \bibinfo{journal}{Phys. Rev. B} \textbf{\bibinfo{volume}{73}},
  \bibinfo{pages}{045322} (\bibinfo{year}{2006}).

\bibitem[{\citenamefont{D'Amico and Vignale}(2000)}]{Amico00}
\bibinfo{author}{\bibfnamefont{I.}~\bibnamefont{D'Amico}} \bibnamefont{and}
  \bibinfo{author}{\bibfnamefont{G.}~\bibnamefont{Vignale}},
  \bibinfo{journal}{Phys. Rev. B} \textbf{\bibinfo{volume}{62}},
  \bibinfo{pages}{4853} (\bibinfo{year}{2000}).

\bibitem[{\citenamefont{Flensberg et~al.}(2001)\citenamefont{Flensberg, Jensen,
  and Mortensen}}]{Flensberg01}
\bibinfo{author}{\bibfnamefont{K.}~\bibnamefont{Flensberg}},
  \bibinfo{author}{\bibfnamefont{T.~S.} \bibnamefont{Jensen}},
  \bibnamefont{and} \bibinfo{author}{\bibfnamefont{N.~A.}
  \bibnamefont{Mortensen}}, \bibinfo{journal}{Phys. Rev. B}
  \textbf{\bibinfo{volume}{64}}, \bibinfo{pages}{245308}
  (\bibinfo{year}{2001}).

\bibitem[{\citenamefont{D'Amico and Vignale}(2001)}]{Amico01}
\bibinfo{author}{\bibfnamefont{I.}~\bibnamefont{D'Amico}} \bibnamefont{and}
  \bibinfo{author}{\bibfnamefont{G.}~\bibnamefont{Vignale}},
  \bibinfo{journal}{Europhysics Lett.} \textbf{\bibinfo{volume}{55}},
  \bibinfo{pages}{566} (\bibinfo{year}{2001}).

\bibitem[{\citenamefont{D'Amico and Vignale}(2002)}]{Amico02}
\bibinfo{author}{\bibfnamefont{I.}~\bibnamefont{D'Amico}} \bibnamefont{and}
  \bibinfo{author}{\bibfnamefont{G.}~\bibnamefont{Vignale}},
  \bibinfo{journal}{Phys. Rev. B} \textbf{\bibinfo{volume}{65}},
  \bibinfo{pages}{85109} (\bibinfo{year}{2002}).

\bibitem[{\citenamefont{D'Amico and Vignale}(2003)}]{Amico03}
\bibinfo{author}{\bibfnamefont{I.}~\bibnamefont{D'Amico}} \bibnamefont{and}
  \bibinfo{author}{\bibfnamefont{G.}~\bibnamefont{Vignale}},
  \bibinfo{journal}{Phys. Rev. B} \textbf{\bibinfo{volume}{68}},
  \bibinfo{pages}{045307} (\bibinfo{year}{2003}).

\bibitem[{\citenamefont{Vignale}(2007)}]{Vignale2007}
\bibinfo{author}{\bibfnamefont{G.}~\bibnamefont{Vignale}},
  \emph{\bibinfo{title}{Manipulating Quantum Coherence in Solid State Systems,
  M. E. Flatte and I. Tifrea (editors)}} (\bibinfo{publisher}{Springer},
  \bibinfo{address}{Berlin}, \bibinfo{year}{2007}).

\bibitem[{\citenamefont{Badalyan et~al.}(2008)\citenamefont{Badalyan, Kim, and
  Vignale}}]{Badalyan08}
\bibinfo{author}{\bibfnamefont{S.~M.} \bibnamefont{Badalyan}},
  \bibinfo{author}{\bibfnamefont{C.~S.} \bibnamefont{Kim}}, \bibnamefont{and}
  \bibinfo{author}{\bibfnamefont{G.}~\bibnamefont{Vignale}},
  \bibinfo{journal}{Phys. Rev. Lett.} \textbf{\bibinfo{volume}{100}},
  \bibinfo{pages}{016603} (\bibinfo{year}{2008}).

\bibitem[{\citenamefont{Weber et~al.}(2005)\citenamefont{Weber, Gedik, Moore,
  Orenstein, Stephens, and Awschalom}}]{Weber05}
\bibinfo{author}{\bibfnamefont{C.~P.} \bibnamefont{Weber}},
  \bibinfo{author}{\bibfnamefont{N.}~\bibnamefont{Gedik}},
  \bibinfo{author}{\bibfnamefont{J.~E.} \bibnamefont{Moore}},
  \bibinfo{author}{\bibfnamefont{J.}~\bibnamefont{Orenstein}},
  \bibinfo{author}{\bibfnamefont{J.}~\bibnamefont{Stephens}}, \bibnamefont{and}
  \bibinfo{author}{\bibfnamefont{D.~D.} \bibnamefont{Awschalom}},
  \bibinfo{journal}{Nature} \textbf{\bibinfo{volume}{437}},
  \bibinfo{pages}{1330} (\bibinfo{year}{2005}).

\bibitem[{\citenamefont{Engel et~al.}(2005)\citenamefont{Engel, Halperin, and
  Rashba}}]{Engel05}
\bibinfo{author}{\bibfnamefont{H.~A.} \bibnamefont{Engel}},
  \bibinfo{author}{\bibfnamefont{B.~I.} \bibnamefont{Halperin}},
  \bibnamefont{and} \bibinfo{author}{\bibfnamefont{E.}~\bibnamefont{Rashba}},
  \bibinfo{journal}{Phys. Rev. Lett.} \textbf{\bibinfo{volume}{95}},
  \bibinfo{pages}{166605} (\bibinfo{year}{2005}).

\bibitem[{\citenamefont{Hankiewicz and Vignale}(2006)}]{Hankiewicz06}
\bibinfo{author}{\bibfnamefont{E.~M.} \bibnamefont{Hankiewicz}}
  \bibnamefont{and} \bibinfo{author}{\bibfnamefont{G.}~\bibnamefont{Vignale}},
  \bibinfo{journal}{Phys. Rev. B} \textbf{\bibinfo{volume}{73}},
  \bibinfo{pages}{115339} (\bibinfo{year}{2006}).

\bibitem[{\citenamefont{Mott and Massey}(1964)}]{Mott}
\bibinfo{author}{\bibfnamefont{N.~F.} \bibnamefont{Mott}} \bibnamefont{and}
  \bibinfo{author}{\bibfnamefont{H.~S.~W.} \bibnamefont{Massey}},
  \emph{\bibinfo{title}{The Theory of Atomic Collisions}}
  (\bibinfo{publisher}{Oxford University Press}, \bibinfo{year}{1964}).

\bibitem[{\citenamefont{Foldy and Wouthuysen}(1950)}]{Foldy50}
\bibinfo{author}{\bibfnamefont{L.~L.} \bibnamefont{Foldy}} \bibnamefont{and}
  \bibinfo{author}{\bibfnamefont{S.~A.} \bibnamefont{Wouthuysen}},
  \bibinfo{journal}{Phys. Rev.} \textbf{\bibinfo{volume}{78}},
  \bibinfo{pages}{29} (\bibinfo{year}{1950}).

\bibitem[{\citenamefont{Winkler}(2003)}]{Winkler2003}
\bibinfo{author}{\bibfnamefont{R.}~\bibnamefont{Winkler}},
  \emph{\bibinfo{title}{Spin-orbit effects in two-dimensional electron and hole
  systems}} (\bibinfo{publisher}{Springer}, \bibinfo{year}{2003}).

\bibitem[{\citenamefont{Landau and Lifshitz}(1964)}]{Landau3}
\bibinfo{author}{\bibfnamefont{L.~D.} \bibnamefont{Landau}} \bibnamefont{and}
  \bibinfo{author}{\bibfnamefont{E.~M.} \bibnamefont{Lifshitz}},
  \emph{\bibinfo{title}{Course of Theoretical Physics, Vol. III.}}
  (\bibinfo{publisher}{Butterworth-Heinemann}, \bibinfo{address}{Oxford},
  \bibinfo{year}{1964}).

\bibitem[{\citenamefont{Kohn and Luttinger}(1957)}]{Kohn58}
\bibinfo{author}{\bibfnamefont{W.}~\bibnamefont{Kohn}} \bibnamefont{and}
  \bibinfo{author}{\bibfnamefont{J.~M.} \bibnamefont{Luttinger}},
  \bibinfo{journal}{Phys. Rev.} \textbf{\bibinfo{volume}{108}},
  \bibinfo{pages}{590} (\bibinfo{year}{1957}).

\bibitem[{\citenamefont{Hankiewicz et~al.}(2006)\citenamefont{Hankiewicz,
  Vignale, and Flatt\'e}}]{HankiewiczPRL06}
\bibinfo{author}{\bibfnamefont{E.~M.} \bibnamefont{Hankiewicz}},
  \bibinfo{author}{\bibfnamefont{G.}~\bibnamefont{Vignale}}, \bibnamefont{and}
  \bibinfo{author}{\bibfnamefont{M.}~\bibnamefont{Flatt\'e}},
  \bibinfo{journal}{Phys. Rev. Lett.} \textbf{\bibinfo{volume}{97}},
  \bibinfo{pages}{266601} (\bibinfo{year}{2006}).

\bibitem[{\citenamefont{Rojo}(1999)}]{Rojo99}
\bibinfo{author}{\bibfnamefont{A.~G.} \bibnamefont{Rojo}}, \bibinfo{journal}{J.
  Phys. Cond. Mat.} \textbf{\bibinfo{volume}{11}}, \bibinfo{pages}{R31}
  (\bibinfo{year}{1999}).

\bibitem[{\citenamefont{Kikkawa et~al.}(1997)\citenamefont{Kikkawa, Smorchkova,
  Samarth, and Awschalom}}]{Kikkawa97}
\bibinfo{author}{\bibfnamefont{J.~M.} \bibnamefont{Kikkawa}},
  \bibinfo{author}{\bibfnamefont{I.~P.} \bibnamefont{Smorchkova}},
  \bibinfo{author}{\bibfnamefont{N.}~\bibnamefont{Samarth}}, \bibnamefont{and}
  \bibinfo{author}{\bibfnamefont{D.~D.} \bibnamefont{Awschalom}},
  \bibinfo{journal}{Science} \textbf{\bibinfo{volume}{277}},
  \bibinfo{pages}{1284} (\bibinfo{year}{1997}).

\bibitem[{\citenamefont{Kikkawa and Awschalom}(1998)}]{Kikkawa98}
\bibinfo{author}{\bibfnamefont{J.~M.} \bibnamefont{Kikkawa}} \bibnamefont{and}
  \bibinfo{author}{\bibfnamefont{D.~D.} \bibnamefont{Awschalom}},
  \bibinfo{journal}{Phys. Rev. Lett.} \textbf{\bibinfo{volume}{90}},
  \bibinfo{pages}{4313} (\bibinfo{year}{1998}).

\bibitem[{\citenamefont{Giuliani and Vignale}(2005)}]{Thebook}
\bibinfo{author}{\bibfnamefont{G.~F.} \bibnamefont{Giuliani}} \bibnamefont{and}
  \bibinfo{author}{\bibfnamefont{G.}~\bibnamefont{Vignale}},
  \emph{\bibinfo{title}{Quantum Theory of the Electron Liquid}}
  (\bibinfo{publisher}{Cambridge University Press}, \bibinfo{address}{UK},
  \bibinfo{year}{2005}).

\bibitem[{\citenamefont{Tse et~al.}(2005)\citenamefont{Tse, Fabian, Žutic, and
  Sarma}}]{Tse05}
\bibinfo{author}{\bibfnamefont{W.-K.} \bibnamefont{Tse}},
  \bibinfo{author}{\bibfnamefont{J.}~\bibnamefont{Fabian}},
  \bibinfo{author}{\bibfnamefont{I.}~\bibnamefont{Žutic}}, \bibnamefont{and}
  \bibinfo{author}{\bibfnamefont{S.~D.} \bibnamefont{Sarma}},
  \bibinfo{journal}{Phys. Rev. B} \textbf{\bibinfo{volume}{72}},
  \bibinfo{pages}{241303} (\bibinfo{year}{2005}).

\bibitem[{\citenamefont{Valet and Fert}(1993)}]{Fert93}
\bibinfo{author}{\bibfnamefont{T.}~\bibnamefont{Valet}} \bibnamefont{and}
  \bibinfo{author}{\bibfnamefont{A.}~\bibnamefont{Fert}},
  \bibinfo{journal}{Phys. Rev. B} \textbf{\bibinfo{volume}{48}},
  \bibinfo{pages}{7099} (\bibinfo{year}{1993}).

\bibitem[{\citenamefont{Winkler}(2000)}]{Winkler00}
\bibinfo{author}{\bibfnamefont{R.}~\bibnamefont{Winkler}},
  \bibinfo{journal}{Phys. Rev. B} \textbf{\bibinfo{volume}{62}},
  \bibinfo{pages}{4245} (\bibinfo{year}{2000}).

\bibitem[{\citenamefont{Shi et~al.}(2006)\citenamefont{Shi, Zhang, Xiao, and
  Niu}}]{Niu06}
\bibinfo{author}{\bibfnamefont{J.}~\bibnamefont{Shi}},
  \bibinfo{author}{\bibfnamefont{P.}~\bibnamefont{Zhang}},
  \bibinfo{author}{\bibfnamefont{D.}~\bibnamefont{Xiao}}, \bibnamefont{and}
  \bibinfo{author}{\bibfnamefont{Q.}~\bibnamefont{Niu}},
  \bibinfo{journal}{Phys. Rev. Lett.} \textbf{\bibinfo{volume}{96}},
  \bibinfo{pages}{076604} (\bibinfo{year}{2006}).

\bibitem[{\citenamefont{Tse and Sarma}(2007)}]{Sarma07}
\bibinfo{author}{\bibfnamefont{W.-K.} \bibnamefont{Tse}} \bibnamefont{and}
  \bibinfo{author}{\bibfnamefont{S.~D.} \bibnamefont{Sarma}},
  \bibinfo{journal}{Phys. Rev. B} \textbf{\bibinfo{volume}{75}},
  \bibinfo{pages}{045333} (\bibinfo{year}{2007}).

\bibitem[{\citenamefont{Weber et~al.}(2007)}]{Weber07}
\bibinfo{author}{\bibfnamefont{C.~P.} \bibnamefont{Weber}}
  \bibnamefont{et~al.}, \bibinfo{journal}{Phys. Rev. Lett.}
  \textbf{\bibinfo{volume}{98}}, \bibinfo{pages}{076604}
  (\bibinfo{year}{2007}).

\bibitem[{\citenamefont{Weng et~al.}(2008)\citenamefont{Weng, Wu, and
  Cui}}]{Weng08}
\bibinfo{author}{\bibfnamefont{M.~Q.} \bibnamefont{Weng}},
  \bibinfo{author}{\bibfnamefont{M.~W.} \bibnamefont{Wu}}, \bibnamefont{and}
  \bibinfo{author}{\bibfnamefont{H.~L.} \bibnamefont{Cui}},
  \bibinfo{journal}{J. App. Phys.} \textbf{\bibinfo{volume}{103}},
  \bibinfo{pages}{063714} (\bibinfo{year}{2008}).

\bibitem[{\citenamefont{Onoda et~al.}(2006)\citenamefont{Onoda, Sugimoto, and
  Nagaosa}}]{Onoda06}
\bibinfo{author}{\bibfnamefont{S.}~\bibnamefont{Onoda}},
  \bibinfo{author}{\bibfnamefont{N.}~\bibnamefont{Sugimoto}}, \bibnamefont{and}
  \bibinfo{author}{\bibfnamefont{N.}~\bibnamefont{Nagaosa}},
  \bibinfo{journal}{Phys. Rev. Lett.} \textbf{\bibinfo{volume}{97}},
  \bibinfo{pages}{126602} (\bibinfo{year}{2006}).

\bibitem[{\citenamefont{Bernevig and Zhang}(2005)}]{Zhang05}
\bibinfo{author}{\bibfnamefont{B.~A.} \bibnamefont{Bernevig}} \bibnamefont{and}
  \bibinfo{author}{\bibfnamefont{S.-C.} \bibnamefont{Zhang}},
  \bibinfo{journal}{Phys. Rev. Lett.} \textbf{\bibinfo{volume}{95}},
  \bibinfo{pages}{016801} (\bibinfo{year}{2005}).

\end{thebibliography}
\end{document}